\documentclass[preprint,11pt, aps,numerical,graphicx,superscriptaddress,longbibliography,showkeys,utf8,onecolumn]{revtex4-2}

\renewcommand{\thesection}{\arabic{section}}
\renewcommand{\thesubsection}{\thesection.\arabic{subsection}}

\makeatletter
\renewcommand{\p@subsection}{}
\renewcommand{\p@subsubsection}{}
\makeatother

\usepackage[margin=3cm]{geometry}
\usepackage[utf8]{inputenc}

\usepackage{tabularx}

\usepackage{amsmath}
\usepackage{amsfonts}
\usepackage{booktabs}
\usepackage{multirow}
\usepackage{natbib}
\usepackage{adjustbox}

\usepackage{hyperref}

\usepackage{graphicx}
\usepackage{dcolumn}
\usepackage{bm}

\usepackage[mathlines]{lineno}

\usepackage[english]{babel}

\usepackage{float}
\usepackage[caption=false]{subfig}

\begin{document}


\title{Mathematical modelling of the distribution of tumor size in patient populations}

\author{Stephan Radonic}
\email[Author to whom correspondence should be addressed. Electronic mail: ]{stephan.radonic@uzh.ch}
\affiliation{Department of Physics, University of Zurich, Zurich, Switzerland}
\affiliation{Division of Radiation Oncology, Small Animal Department, Vetsuisse Faculty, University of Zurich, Zurich, Switzerland}

\author{Jürgen Besserer}
\affiliation{Radiotherapy Hirslanden AG, Rain 34 , Aarau, Switzerland}

\author{Jessica Kneubühl}
\affiliation{Department of Physics, University of Zurich, Zurich, Switzerland}

\author{Valeria Meier}
\affiliation{Department of Physics, University of Zurich, Zurich, Switzerland}
\affiliation{Division of Radiation Oncology, Small Animal Department, Vetsuisse Faculty, University of Zurich, Zurich, Switzerland}

\author{Uwe Schneider}
\affiliation{Department of Physics, University of Zurich, Zurich, Switzerland}
\affiliation{Radiotherapy Hirslanden AG, Rain 34 , Aarau, Switzerland}

\date{\today}

\begin{abstract}
In radiation therapy tumor size, and thus also volume, has a significant impact on the local control of tumors. Moreover, tumor volume is a significant prognostic factor for modelling and predicting therapeutic outcomes in cancer treatment.  In research, the distribution of tumor volumes in patient populations has so far remained widely unexplored. In this work, the frequency distributions of maximum diameter of tumors for various types of cancer was studied based on SEER data and it was explored if they can be modelled using Weibull distributions. Further, actual volume data were obtained directly from computer tomography (CT) datasets and the frequency distributions of tumor volumes and maximum diameters were explored and a link between them was found. In cancer research, tumors are often modelled as ellipsoids. In order to verify the appropriateness of using ellipsoids as a model, to the obtained three-dimensional data, ellipsoids were fitted and the resulting volumes and diameters analysed.  Finally, NSCLC tumor diameters were Monte Carlo simulated using tumor growth and incidence models. A comparison of the simulated tumor diameter distributions with observed SEER data yielded the determination of tumor growth rates.
\end{abstract}

\maketitle

\section{Introduction}
Tumor size is an important parameter in various aspects of cancer research. The tumor size, its shape and its volume are important variables for tumor growth models. Moreover, it is a significant prognostic factor for modelling and predicting therapeutic outcomes in cancer treatment \citep{Dubben1998}. In radiation therapy, tumor size and thus also volume has a significant impact on the local control of tumors. The impact of tumor volume on tumor control has been the subject of various studies \citep{Buffa2000, Brenner1993, Fenwick1998, Bentzen1996, Webb1994}.

The tumor control probability (TCP) model established in radiation therapy is as follows:
\begin{equation}
TCP = e^{-N\cdot S}
\end{equation}
where $N$ is the initial number of clonogenic cells and $S$ is the surviving fraction after irradiation. Thereby $S$ is a function of the radiation dose applied. The most established cell survival function is the linear quadratic (LQ) model. The cell number $N$ can be expressed as the product $N=\rho\cdot V$ of a clonogenic cell density $\rho$ and the tumor volume $V$. In general $\rho$ could also be assumed to be inhomogeneous within the tumor volume and to even have a functional dependence on the volume size $\rho(V)$. In any case the cell number $N$ is proportional to the volume $V$. Thus it becomes obvious that tumor volume is a crucial parameter for TCP modelling.  In order to obtain the required model parameters the model has to be fitted to clinical tumor control (or disease specific survival) data, which always stem from a heterogenous patient (cancer) population. The Poisson model considers a perfectly homogeneous patient population and yields TCP curves too steep to accurately match cliniclly observed control data \citep{Webb1993}. Hence, population TCP models are studied. When investigating population averaged TCP models, variables which are assumed to differ within a patient population are identified and used to obtain a population TCP model from the Poisson TCP model. Typically such variables are the radio-sensitivity of the cancer cells, the cell density and the tumor volume \citep{Webb1993, Schinkel2007, Carlone2006, Buffa2000, Keall2007, Roberts1998}. Such models provide a better fit to clinical data. However,  they carry disadvantages such as introducing additional parameters and rarely having a closed form. Often empirical models such as the \emph{logit} are used, they provide a good fit to clinical data while only using a small number of parameters \citep{Okunieff1995}. In \citet{SCHNEIDER2022143, Radonic2021} a closed form population TCP model was obtained which characteristically matches the logistic model used by \citet{Okunieff1995}. This was done by assuming that the tumor volumes in the patient population are exponentially distributed. In \citet{Radonic2021} a sample of gross tumor volume (GTV) sizes of canine patients which had undergone radiation therapy for brain tumor was analysed. It was found that a slightly modified exponential distribution, where a limited detection rate for very small tumors is assumed, matches the frequency distribution of the observed clinical volume data. 

In research, the distribution of tumor volumes in patient populations has so far remained widely unexplored. The literature search for published data sets of tumor volumes, also yielded very few results. Most publications contain only the largest diameter of the tumors. The Surveillence , Epidemiology and End Results (SEER)\citep{sd21} database of the U.S. National Cancer Institute contains a vast amount of tumor patient data for various cancer diagnoses, which also includes the largest tumor diameter, also referred to as "tumor size". In \citet{RAMACHANDRAN2021569}, SEER data for prostate cancer were analysed, it was discussed that the frequency distribution of tumor sizes might follow a Weibull distribution. To our knowledge there is no mechanistic explanation for this. Moreover, the largest tumor diameter for itself does not allow to directly infer the tumor volume, since the shape is unknown. In cancer research often spherical, cylindrical or ellipsoidal shapes are assumed \citep{Heuser1979}. \\

In literature various tumor growth model have been proposed and investigated. \citet{Talkington2015} investigated different tumor growth models and inferred parameters of in vivo growth for different types of cancer. In their work the parameters were derived from an extrapolation of two data points.

The goals of this work were to 1) study the frequency distributions of maximum diameter of tumors for various types of cancer based on SEER data and particularly to explore if they follow Weibull distributions, 2) obtain actual volume data directly from computer tomography (CT) datasets and explore frequency distributions of volumes and maximum diameters 3) in order to verify how well ellipsoids resemble the actual tumor shapes, fit ellipsoids to the three dimensional tumor data from CTs, 4) try to establish a link between the frequency distributions of tumor volumes and diameters and 5) perform a simulation of tumor volumes using a tumor growth model and compare the results with the diameter and volume distribution data from SEER and CTs.

\section{Materials and methods}

\subsection{SEER data}

In order to obtain tumor size (diameter) data, we have used data from the Surveillence , Epidemiology and End Results (SEER) database\citep{sd21} of the U.S. National Cancer Institute. 
We have considered \emph{breast cancer},  \emph{non-small-cell lung cancer (NSCLC)}, \emph{meningioma}, \emph{nasopharyngal cancer} and \emph{liver cancer}. Generally, the SEER\cite{sd21} database provides data since 1975. Per SEER guidelines the tumor size shall be measured as the largest diameter or dimension of the tumor.  The way how the tumor size has been coded in SEER has changed over time. For data after 1988 the SEER tumor size coding is in principle laid out to provide millimetre precision, while between 1975 to 1988 data were categorized with a much worse precision in the order of one centimetre.  Because of this and due to very high number of available cases/patients recorded in SEER, for this work only data after 1988 was considered. 
The data were queried using SEERStat\citep{seer}.
In Table~\ref{table1:overview_seer} an overview of the data used is provided. The data obtained from the SEER queries were filtered such that only the relevant tumor diameter ranges are included. For breast cancer the data were constrained to only female patients aged over 15 years. 
\begin{table}[h!]
\begin{tabular}{@{}lllll@{}}
\toprule
\textbf{cancer type} & \textbf{diameter range {[}cm{]}} & \textbf{\# tumor cases} & \textbf{\% of dataset} & other constraints             \\ \midrule
Breast               & 0 - 10                           & 477471              & 99.21                  & female, age \textgreater 15 y \\
NSCLC                & 0 - 15                           & 193580              & 99.76                  & -                             \\
Meningioma           & 0 - 15                           & 402                 & 99.75                  & -                             \\ 
Nasopharynx          & 0 - 11                           & 2413                & 99.38                  & -                             \\
Liver                & 0 - 25                           & 31168               & 99.76                  & -                      \\
\bottomrule      
\end{tabular}
\caption{\textbf{Overview of analysed SEER data}}

\begin{flushleft} \vspace{0.1cm}
\textbf{diameter range}: the tumor diameter range to which the analysis and fitting procedure was constrained to; \textbf{\# tumor cases}: the number of cases in the selected diameter range; \textbf{\% of dataset}: proportion of tumors from total data set from the respective query which lie within the selected diameter range
\end{flushleft}

\label{table1:overview_seer}

\end{table}
The fitting of the Weibull distributions to the data was done using Python and the \emph{lmfit} library.

\subsection{Three dimensional data from computer tomography}

The cancer imaging archive (TCIA) \citep{cvsfkkmpmptp13} contains various collections of imaging data (CT, MRI, PET...) of cancer patients.  For patients treated with external beam radiation therapy (RT), the standard procedure involves doing a planning CT of the treatment region. The planning CT is registered with the other diagnostic imaging data which have been made, such as MRI. With the entirety of the information available from imaging, the radiation oncologist draws the contours of the tumor, also called the gross tumor volume (GTV) on the planning CT. In DICOM format \emph{RTSTRUCT}, the contoured structures are sets of slices. Each slice holds an ordered sequence of points (x,y,z - triplets) defining a contour. The Imebra DICOM library \citep{imebraImebraSDK} was used to read the contour data from the DICOM files.

From TCIA we obtained a collection of CT datasets of patients which had undergone radiation therapy for NSCLC \citep{awvehpgcbmhrhrldqgl19}. The collection contained data from 422 NSCLC patients. For these patients, pretreatment CT scans along with manual delineations of the GTVs were available. We have considered only cases where solely a single GTV was present which was the case for 176 patients. 
\\
Further, CT datasets of 96 canine patients which had undergone RT for meningioma at our institution in the time period between xx and xx, were obtained from the Varian ARIA database using Varian Eclipse Scripting API. 

\subsection{Estimation of volume and diameter}
\label{sec:estvol}
Using the 3D contour data, the volume of the GTVs was calculated using a custom algorithm which was implemented in C++. The algorithm calculates the contoured area for each slice. The area is then multiplied by the CT slice thickness. For the canine meningioma data, the volume was also calculated using a function provided by Varian ESAPI. This was done in order to verify and asses the precision of our algorithm.

The contour data were also used to obtain the largest distance between two contour (surface) points within the GTV structure. This is done by computing 
\begin{equation}
\max \left(||\mathbf{r}_i-\mathbf{r}_j||_{\forall i,j \in Surf_{GTV} | i\neq j}\right).
\end{equation}
The algorithm was verified by randomly selecting a few patients, for which the largest diameter was measured manually using Varian Eclipse. We believe the obtained metric to be comparable to the tumor size (largest diameter) as coded in SEER. 

\subsection{Weibull distribution}

The Weibull distribution is a continuous probability distribution, its probability density function (PDF) is given by 
\begin{equation}
f(x;\lambda,k) = \frac{k}{\lambda}\left(\frac{x}{\lambda}\right)^{k-1} e^{-\left(\frac{x}{\lambda}\right)^k}~,~x\geq0.
\label{eq:weibull}
\end{equation}

\subsection{Frequency distribution of tumor volumes}
An interesting property of the Weibull distribution is that if a random variable X is exponentially distributed as 
\begin{equation}
f_\lambda(x)=\frac{1}{\lambda}e^{-\frac{x}{\lambda}},
\label{eq:expon}
\end{equation},
the random variable $Y:=X^{1/k}$, $(k>0)$ follows a Weibull Distribution with $Wei(\lambda^{1/k},k)$.
In research, the shape of tumors is often modelled as a spheroid, cylinder or ellipsoid. Particularly the ellipsoid is very versatile as it covers the special case of a spheroid and it can, up to a certain extent, also fairly resemble cylindrical shapes. If an ellipsoidal shape is assumed, the volume of a tumor is then given by 
\begin{equation}
V = \frac{\pi}{6}d_a d_b d_c
\end{equation}
where $d_a$, $d_b$ and $d_c$ are the diameters of the three ellipsoid axes. In order to establish a relation between the tumor volume $V$ and its largest diameter $d_{max}$, it is assumed that 
\begin{equation}
V = \frac{\pi}{6}d_a d_b d_c \overset{!}{=}  \frac{\pi}{6}\mu^{3-m}\cdot(d_{max})^m.
\label{eq:elips_vol}
\end{equation}
where $\mu^{3-m}$ is a scaling factor which ensures the correctness of the volume unit and scale invariance. It has length units (e.g. cm). The volume always has to have a dimension of length cubed (e.g. cm$^3$).
The exponent $m$ can now without loss of generality be expressed as 
\begin{equation}
m=\frac{k}{\gamma},
\label{eq:m_k_gamma}
\end{equation} where $k$ is the parameter of the Weibull distribution as in Eq.~\ref{eq:weibull}, and $\gamma \in \mathbb{R}:\gamma=\frac{k}{m}$. The variable $\gamma$ has to be introduced as the Weibull distribution of the maximal diameters contains no information about the eccentricity of the ellipsoids. 
If $X$ has a probability distribution $f_X(x)$, and a variable $Y$  is a function $y = g(x)$ of $X$, then the probability distribution of $Y$ is given by
\begin{equation}
f_Y(y) = f_X(g^{-1}(y))\frac{dg^{-1}(y)}{dy}
\label{eq:probtrafo}
\end{equation}.
Assuming that $d_{max}$ is distributed according to Eq.~\ref{eq:weibull}, and that the function $y=g(x)$ is given by relation between $V$ and $d_{max}$ is $V=\frac{\pi}{6}\mu^{3-\frac{k}{\gamma}}\cdot(d_{max})^{\frac{k}{\gamma}}$ from Eq.~\ref{eq:elips_vol}, according to Eq.~\ref{eq:probtrafo} the PDF of the tumor volumes is then given by
\begin{equation}
f(V;\lambda,\gamma,k,s)=\frac{\gamma  \left(\frac{6}{\pi }\right)^k e^{-\left(\frac{6}{\pi }\right)^k \left(\frac{\mu ^{1-\frac{3 \gamma }{k}} V^{\frac{\gamma}{k}}}{\lambda }\right)^k} \left(\frac{\mu ^{1-\frac{3 \gamma }{k}} V^{\frac{\gamma}{k}}}{\lambda }\right)^k}{V}.
\label{eq:voldistr}
\end{equation}
By comparing Eq.~\ref{eq:voldistr} with a Weibull distribution
\begin{equation}
f(V;\lambda_V,k_V) = \frac{k_V}{\lambda_V}\left(\frac{V}{\lambda_V}\right)^{k_V-1} e^{-\left(\frac{V}{\lambda_V}\right)^{k_V}}~,~V\geq0
\label{eq:weibull_e}
\end{equation}
it can be observed that Eq.~\ref{eq:voldistr} is also a Weibull distribution with 
\begin{equation}
k_V=\gamma
\label{eq:k_vol}
\end{equation} and
\begin{equation}
\lambda_V=\frac{\left(\frac{\pi }{6}\right)^{k/\gamma } \lambda ^{k/\gamma }}{\mu ^{\frac{k}{\gamma }-3}}
\label{eq:lambda_vol}
\end{equation}
Thus if $(\gamma \text{ or } m)$ and $\mu$ is known the distribution of volumes can be calculated from a Weibull fit to the distribution of maximum diameters. It can be noticed that the parameter $\mu$ only influences (scales) $\lambda_V$.   For $k_V=\gamma=1$, Eq.~\ref{eq:voldistr} transforms to the exponential distribution.  \\
\\
In this work, the parameter $\mu$ was set to be $\mu=1$~cm. This was done as such, because when fitting Eq. ~\ref{eq:elips_vol} to corresponding data, it was found that the fit does not improve much when both parameters $\mu$ and $m$ are fitted, compared to when only $m$ is fitted while $\mu$ is kept fixed. Also it was found that the parameters $\mu$ and $m$ are highly correlated. The unit of $\mu$ was set to cm, as the equation was fitted to data in centimetre units.

\subsection{Fitting ellipsoids to tumor structures}
The ellipsoid is often used as a model for tumor shapes in literature. We used the extracted GTV contour data as point clouds to which ellipsoids were fitted. For this purpose the procedure presented in \citet{Ying2012} was used. We have implemented the procedure in C++ using the Eigen library \cite{eigenweb}. A detailed description of the methodology can be found in appendix A.

\subsection{Analysis of data from computer tomography imaging}

We performed an analysis of the obtained volume and maximal tumor diameter data and compared it with the obtained ellipsoids. Both the volumes and maximal diameter of the actual GTVs and of the fitted ellipsoids were histogramed. The number of bins $N_{bins}$ was chosen such that $N_{bins}<\sqrt{N}$, where $N$ is the sample size. We calculated errorbars on the histograms as
\begin{equation}
\sigma_i = m^n_i \cdot\frac{\sqrt{m_i}}{m_i},
\label{eq:histogram_err}
\end{equation}
where $m^n_i$ is the value of the bin $i$ in the normalized histogram and $m_i$ is the number of data points in bin $i$.
Weibull distributions were fitted to the so obtained data points. From the parameters $\lambda$ and $k$ obtained from Weibull fits to diameter data, and the parameter $m$ determined from fitting Equation~\ref{eq:elips_vol} to the diameter and volume data, Equations~\ref{eq:k_vol}, \ref{eq:m_k_gamma} and \ref{eq:lambda_vol}  were used to calculate the parameters  $\lambda_V$ and $k_V$ for a Weibull distribution for the volumes (as in Eq.~\ref{eq:weibull_e}). The errors on these variables were calculated using propagation of uncertainty while potential correlations between the variables were neglected.
For comparison with \citet{Radonic2021}, to the histograms of the tumors volumes of the canine meningioma, additionally the exponential distribution with the mean volume of the sample $\overline{V}$ was plotted, as well as the exponential distribution with an assumed detection limitation
\begin{equation}
f(V)=\frac{\overline{V}+V_C}{\overline{V}^2}\left(1-\exp\left(-\frac{V}{V_C}\right)\right)\exp\left(-\frac{V,}{\overline{V}}\right)
\end{equation} where $\overline{V}$ is the mean volume and $V_C$ is a surrogate measure for detection limitation of small volumes \citep{Radonic2021}. For this $\overline{V}$ was fixed to the mean of the sample, while $V_C$ was determined by fitting. 

For both the NSCLC and canine meningioma data, the analysis was constrained to tumors where the difference between the volume of the fitted ellipsoid and the volume of the corresponding GTV was not larger than $35\% $. For the analysis data of canine meningioma, the data was filtered such that only tumors with a volume below 10 cm$^3$ were analysed. This was done as there were only 2 data points with volumes above 10 cm$^3$. For human NSCLC, the data was not filtered further and the entire volume range was considered. In some cases, either the single GTV contained multiple disjoint structures, other artifacts were present or the fitting procedure did not produce an ellipsoidal solution. For NSCLC this was the case for 9 patients, for canine meningioma for 3 patients. This left us with 167 data points for human NSCLC and 93 data points for canine menigioma. After applying the aforementioned constraint on the volume difference, for NSCLC 150 (90\%), and for canine meningioma 67 (72\%) data points were left for analysis.

\subsection{Tumor growth modelling}
A combined simulation and fitting procedure was used in order to replicate the NSCLC tumor diameter distribution from SEER.
For this purpose, we devised a Monte Carlo scheme in order to simulate tumor diameter and volume distribution observed in patient population.
\citet{Shuryak2009}\citep{Shuryak2009_2} modelled the mean expected number of new malignant cell incidence in the population as 
\begin{equation}
A(t) = \frac{a}{b}\left(\exp(b\cdot t)-1\right)\exp(-ct^2),
\label{eq:incid_growth}
\end{equation}
where $t$ is the age at induction. It is used as an approximation for the cancer incidence hazard function after some latency period of $L$ years \citep{Shuryak2009, Shuryak2009_2}. Shuryak et al. \citet{Shuryak2009, Shuryak2009_2} fitted the model to background cancer incidence data from SEER database for various cancer types and obtained values for the parameters $a,b$ and $c$. When normalized such that $\int_{-\infty}^{\infty} A(t) dt \overset{!}{=} 1$, Eq.~\ref{eq:incid_growth} is then transformed to
\begin{equation}
\frac{\sqrt{c}}{(\exp(\frac{b^2}{4c})-1)\sqrt{\pi}}\left(\exp(b\cdot t)-1\right)\exp(-ct^2).
\label{eq:incid_growth_norm}
\end{equation}

As a model for tumor growth we have used the power law which generalizes the exponential growth model \cite{Talkington2015}
\begin{equation}
\frac{dV}{dt}=rV(t)^\alpha
\end{equation}
For $\alpha=1$, the model is equivalent to the exponential growth. For $\alpha<1$ the solution, as described in \citep{Talkington2015}, is 
\begin{equation}
V(t)=\left(V_0^{1-\alpha}+(1-\alpha) r t\right)^{1 /(1-\alpha)}.
\label{eq:power_growth}
\end{equation}
\citet{Mayneord1932OnAL} suggested $\alpha=2/3$, which corresponds to the assumption that the tumor only grows at its surface \citep{Talkington2015}. 
\newline
\newline
Our simulation is then set up as follows: 
\begin{enumerate}
\item We assume the geometrical shapes of the tumors to be ellipsoids, particularly we assume two semi-diameters of the ellipsoids to be equal $d_b=d_c=d_{min}$. As in Eq.~\ref{eq:elips_vol} the volume of the ellipsoid is given by
\begin{equation}
V =\frac{\pi}{6}d_{max}(d_{min})^2 \overset{!}{=}  \frac{\pi}{6}\mu^{3-m}(d_{max})^m,
\label{eq:vol_el}
\end{equation}
the large and the small diameters are then related as
\begin{equation}
d_{min} = \sqrt{\mu^{3-m}(d_{max})^{m-1}}
\end{equation}
As was stated previously, it was assumed that $\mu=1$ cm. 
\item First we calculate the mean latency time $L$ using Eq.~\ref{eq:power_growth}. For this an initial volume $V_0$ of $10^{-9}$ cm$^3$ is assumed, which corresponds to a single malignant cell. From the tumor diameter data from SEER, for each tumor $i$ the volume $V_i$ is calculated with Eq.~\ref{eq:vol_el}. Eq.~\ref{eq:power_growth} is then solved for $t$, the result being the latency time $L_i$. From this, the mean latency time $L$ is calculated.

\item We assume that the tumors are observed at a fixed age $age_{det}$, which we fixed to $age_{det}=70$. Accordingly, NSCLC maximum diameter tumor data from SEER were filtered to cases, where the age at diagnosis was equal to 70 years. There were 6917 cases in the filtered SEER data set. In the simulation, 10'000 virtual tumors are sampled. In a single simulation iteration we first sample the age at which a malignant cell is induced $age_{ind}$ from Eq.~\ref{eq:incid_growth_norm} using $L$. This enables us to deduce the growth time $t_{grow}$. If $t_{grow}>0$ the resulting tumor volume is calculated according to Eq.~\ref{eq:power_growth}. Further, it is assumed that not all tumors are detected, this is modelled by assuming a volume dependent detection rate 
\begin{equation}
D(V)=1-\exp(-V/V_C),
\label{eq:detectmodel}
\end{equation}
where $V_C$ is a surrogate measure for limited detection capability. We then sample a uniformly distributed random real number $d_r$ in the interval $[0,1]$, if $d_r<D(V)$ the tumor is detected, otherwise not. Finally for the detected tumors, we calculate $d_{max}$ and $d_{min}$ from the volume $V$ as explained in step 1. 
\end{enumerate} 
The simulation requires the tumor growth rate $r$, the parameter for detection rate $V_C$ and the geometric factor $m$ as input parameters. 
\newline
\newline
A simulation yields a set of tumor diameters. We can combine the simulation with a fitting procedure. Normalized histograms with the exact same binning are computed both from diameter data from SEER, as well as from the simulation. We can then compute a chi-square by summing the difference over all histogram bins
\begin{equation}
\chi^2 = \sum_i^{N_{bins}} (H^{sim}_i - H^{SEER}_i)^2,
\end{equation}
where $H^{sim}$ depends on the input parameter set $\mathbf{p}$. The "optimal" parameters are then found by minimizing chi-square. \newline
\newline
We performed this procedure for NSCLC. The parameters for the malignant cell incidence distribution (Eq.~\ref{eq:incid_growth_norm}) were taken from \citet{Shuryak2009_2}. In the simulation, regarding the parameter $m$, two different variants were tried: One was to set $m = \mu_m$ constant for all tumors, as determined from the analysis of three dimensional NSCLC data. The second variant was to assume $m$ to be normally distributed, with the parameters $\mu_m$ and $\sigma_m$ determined from the analysis of three dimensional NSCLC data.

\section{Results}

\subsection{Weibull fitting with SEER data}
\begin{figure}[!h]
 \centering
\subfloat[\label{weibull_seer:a}]{%
\includegraphics[width=0.49\textwidth]{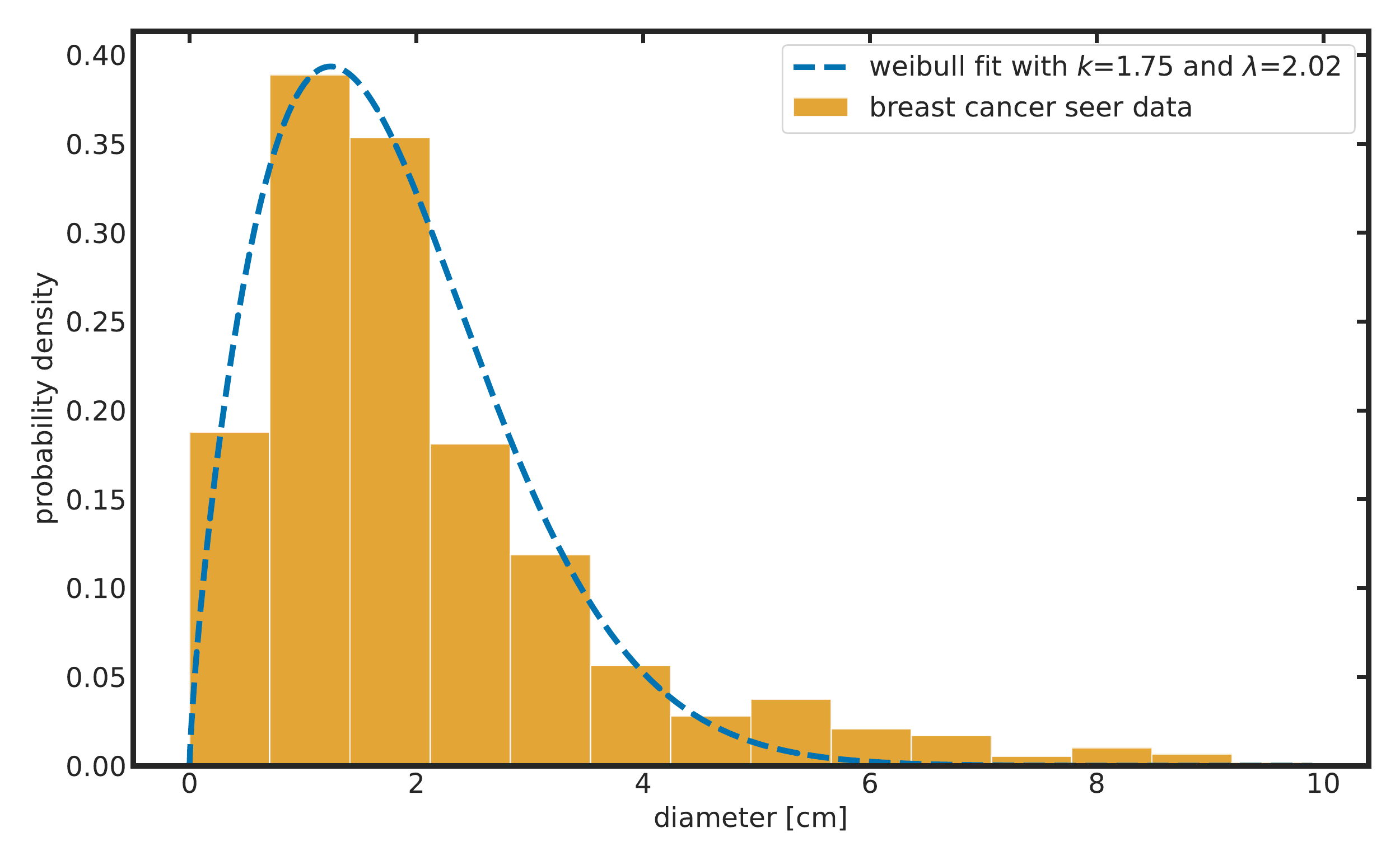}%
}\hfill
\subfloat[\label{weibull_seer:b}]{%
\includegraphics[width=0.49\textwidth]{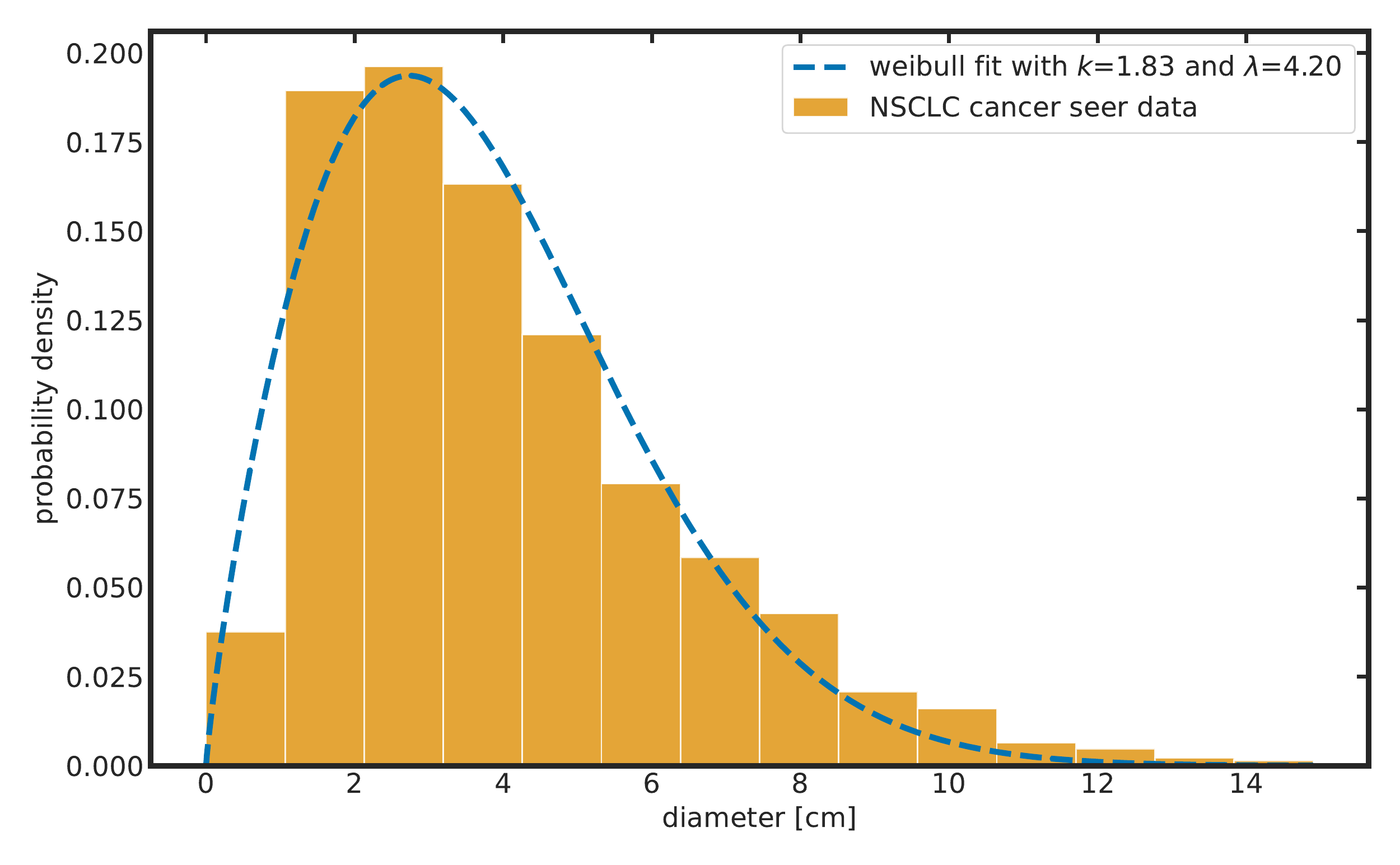}%
}\hfill
\subfloat[\label{weibull_seer:c}]{%
\includegraphics[width=0.49\textwidth]{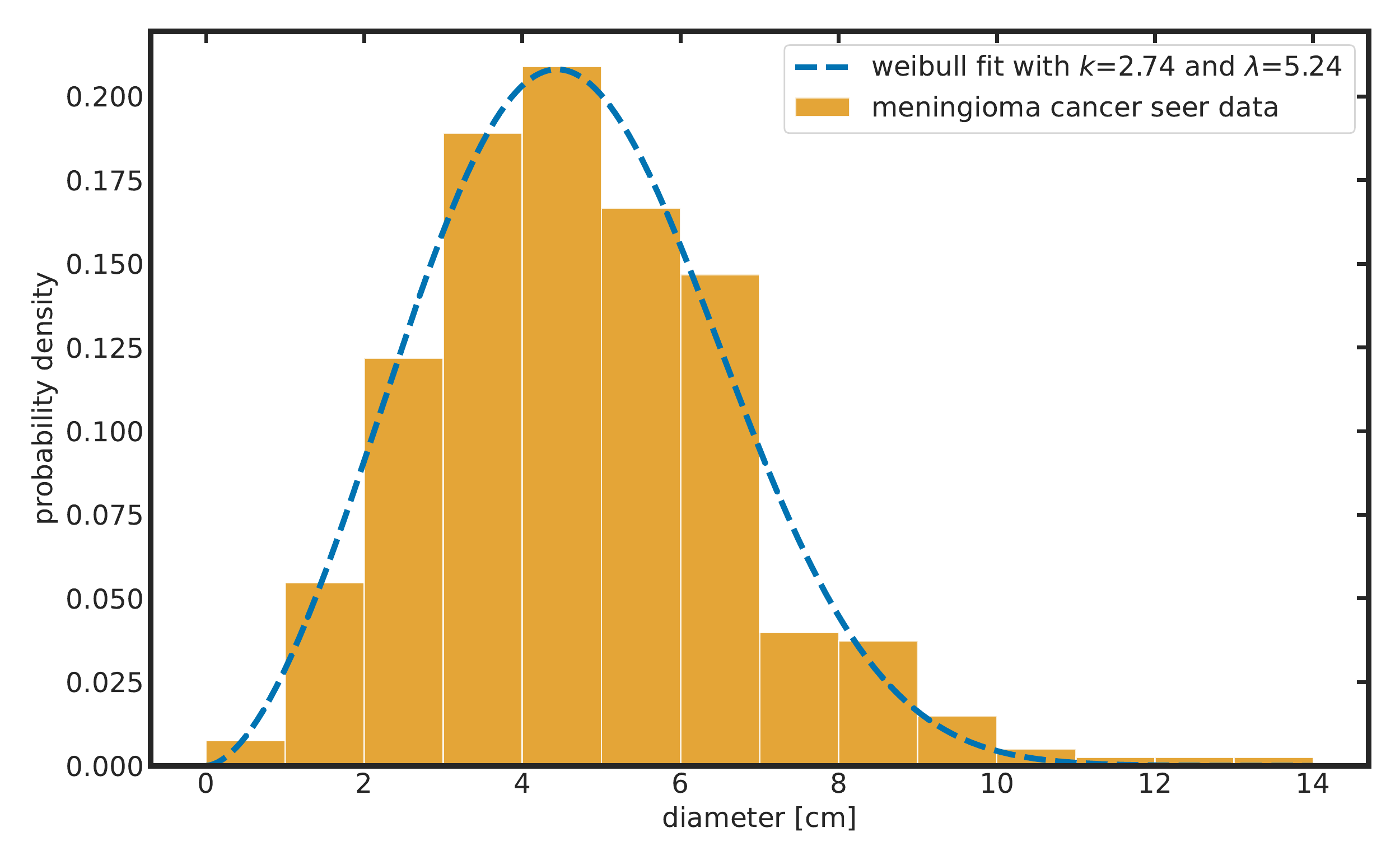}%
}\hfill
\subfloat[\label{weibull_seer:d}]{%
\includegraphics[width=0.49\textwidth]{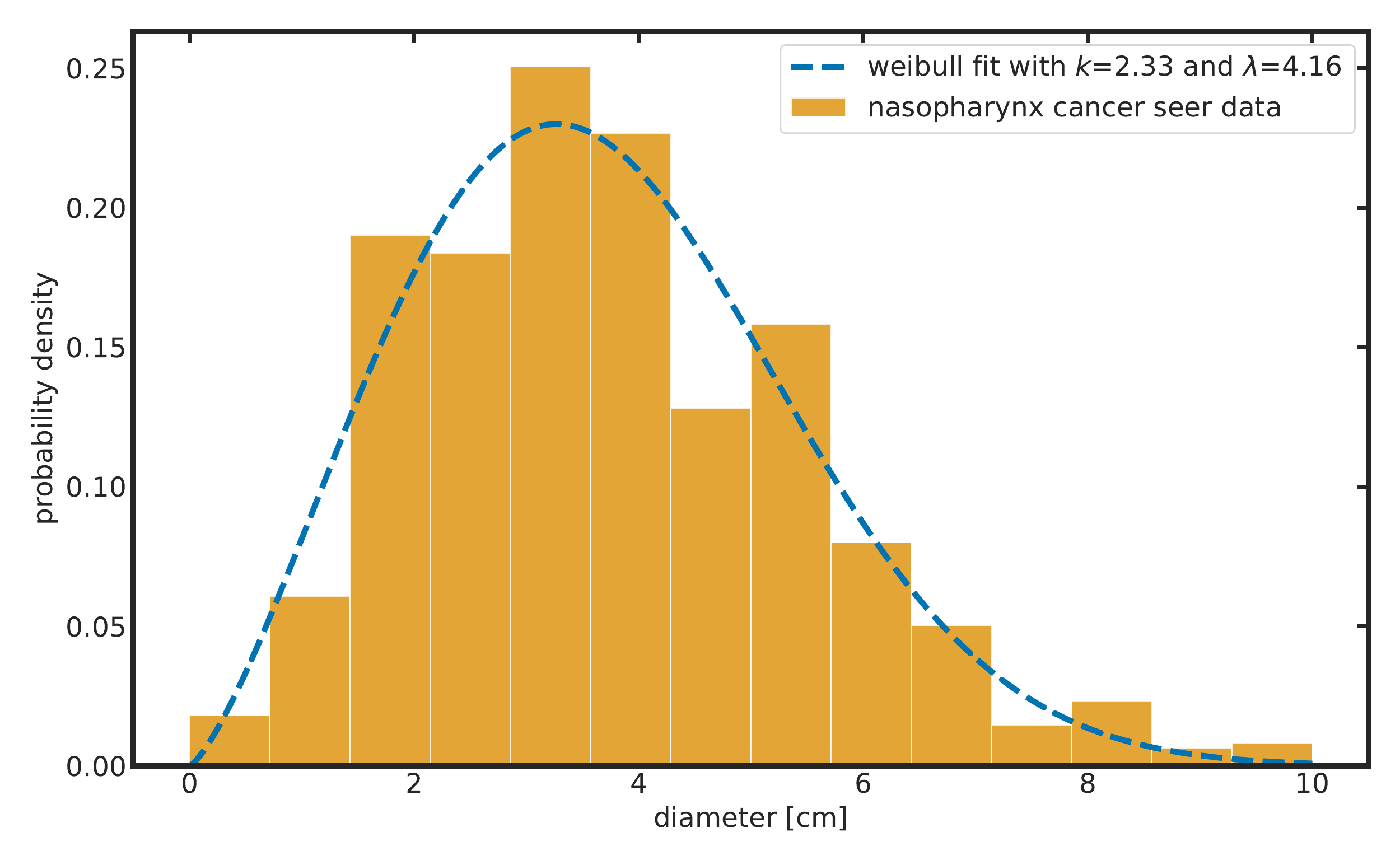}%
}\hfill
\subfloat[\label{weibull_seer:e}]{%
\includegraphics[width=0.49\textwidth]{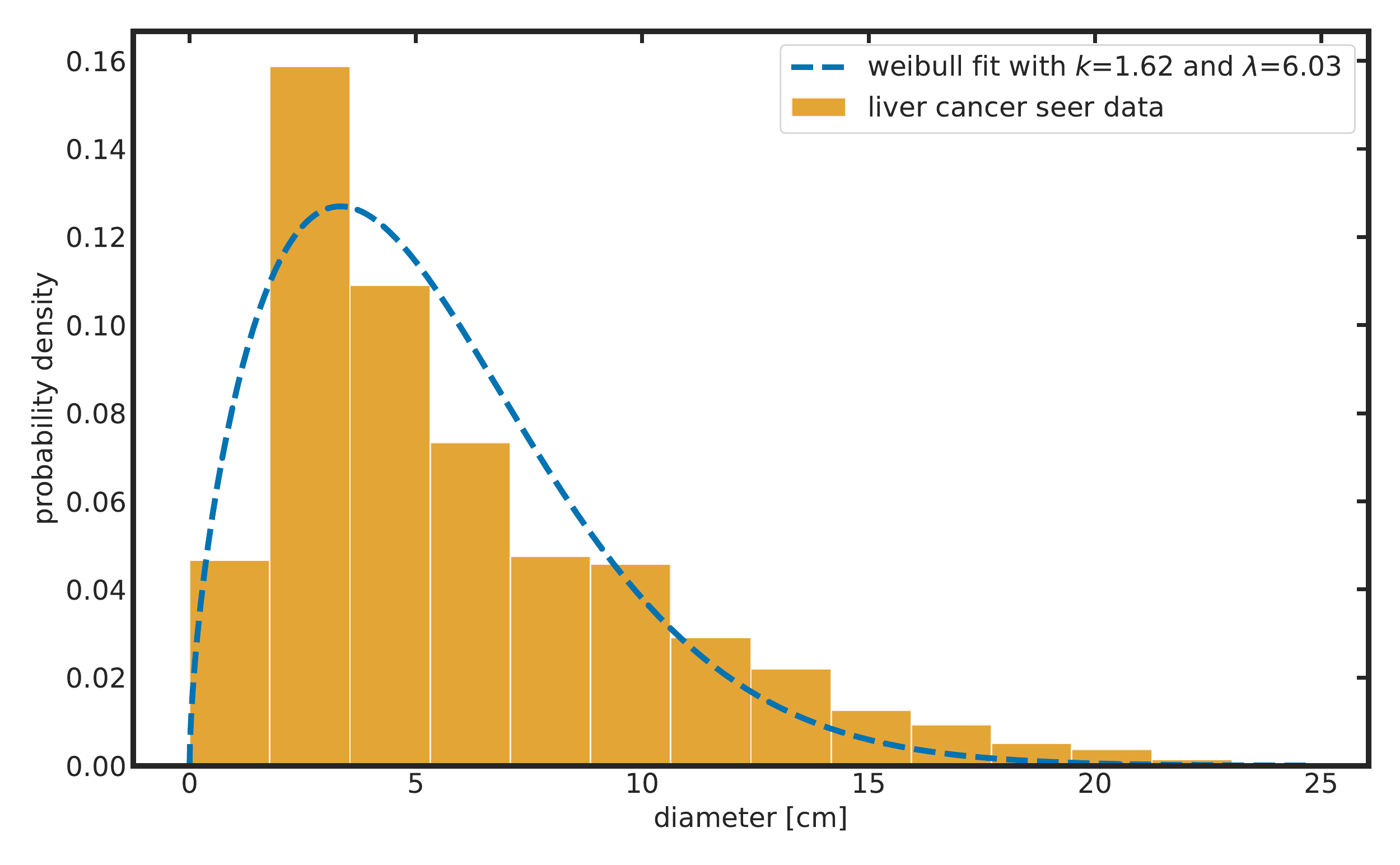}%
}
\caption{\textbf{Tumor size distributions of various cancer types with corresponding Weibull distribution fits}
A: Breast; B: NSCLC; C: Meningioma; D: Nasopharynx; E: Liver}
\label{fig:weibull_seer}
\end{figure}

In Figure~\ref{fig:weibull_seer} the histograms of the tumor size data from the SEER database with the corresponding Weibull distribution fits are shown. Table~\ref{table:results_seer} provides an overview of these results. 

\clearpage
\newgeometry{right=1cm,left=1cm}

\clearpage

\restoregeometry
\begin{table}[!h]

\begin{tabular}{@{}lclclccc@{}}
\toprule
\textbf{cancer type} &
  k &
   &
  \textbf{$\lambda$} &
   &
  \textbf{$\chi^2$} &
  \textbf{\begin{tabular}[c]{@{}c@{}}mean \\ diameter {[}cm{]}\end{tabular}} &
  \textbf{\begin{tabular}[c]{@{}c@{}}median \\ diameter {[}cm{]}\end{tabular}} \\ \midrule
Breast      & $1.75 \pm 0.07$ &  & $2.02 \pm 0.07$ &  & 0.0069 & 2.08 & 1.6 \\
NSCLC       & $1.83 \pm 0.09$ &  & $4.20 \pm 0.15$ &  & 0.0031 & 4.04 & 3.5 \\
Meningioma  & $2.74 \pm 0.1$  &  & $5.24 \pm 0.09$ &  & 0.0016 & 4.58 & 4.3 \\
Nasopharynx & $2.33 \pm 0.13$ &  & $4.15 \pm 0.13$ &  & 0.0067 & 3.75 & 3.5 \\
Liver       & $1.62 \pm 0.13$ &  & $6.03 \pm 0.42$ &  & 0.0035 & 6.05 & 4.9 \\ \bottomrule
\end{tabular}
\caption{\textbf{Resulting parameters from fitting Weibull distributions to SEER data}}
\label{table:results_seer}

\end{table}

\subsection{Analysis of three dimensional data from contoured computer tomography images}
\begin{figure}[!h]
 \centering
\subfloat[\label{vols_ellips_vs_vol_act:a}]{%
\includegraphics[height=0.33\textheight]{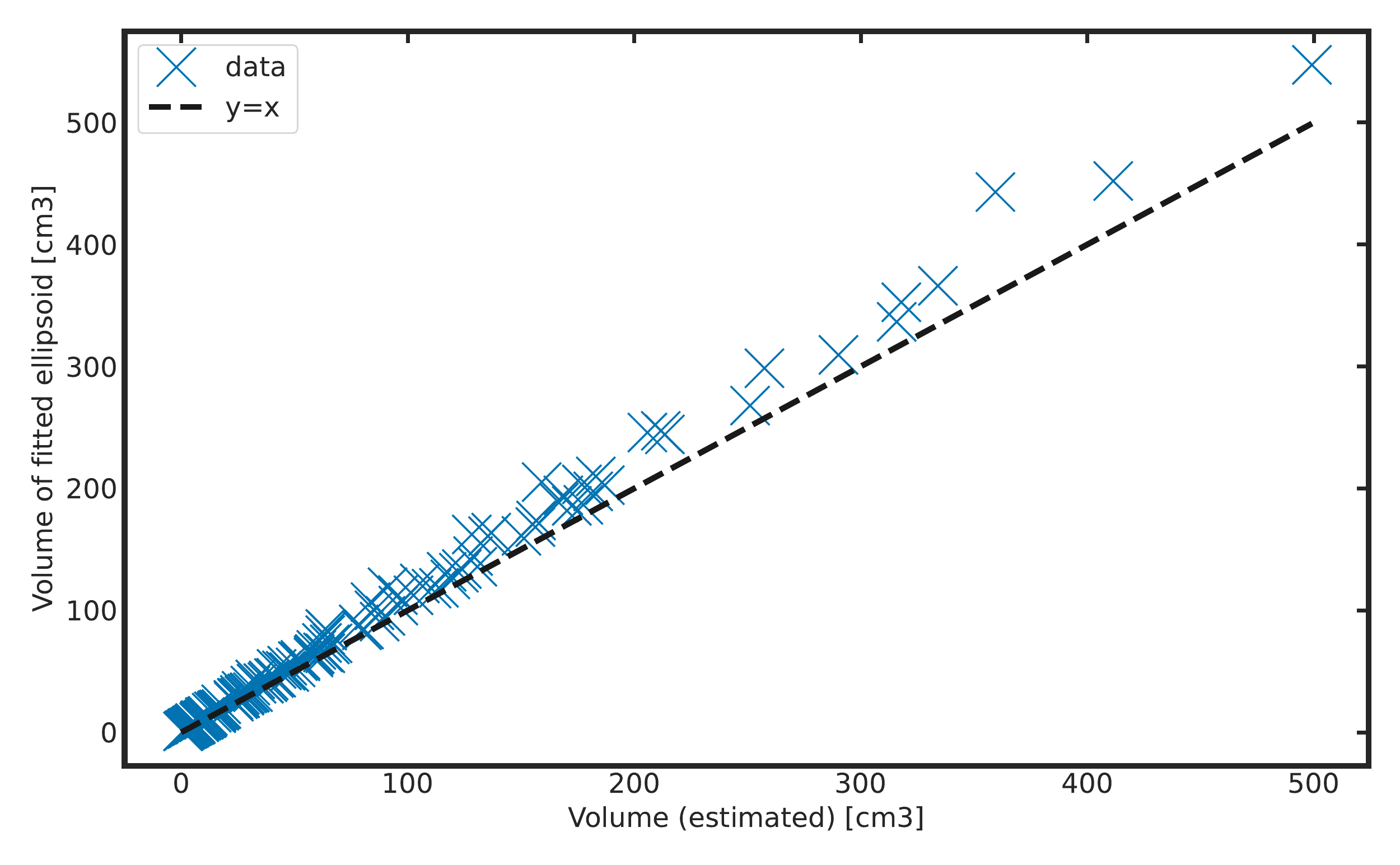}%
}\hfill
\subfloat[\label{vols_ellips_vs_vol_act:b}]{%
\includegraphics[height=0.33\textheight]{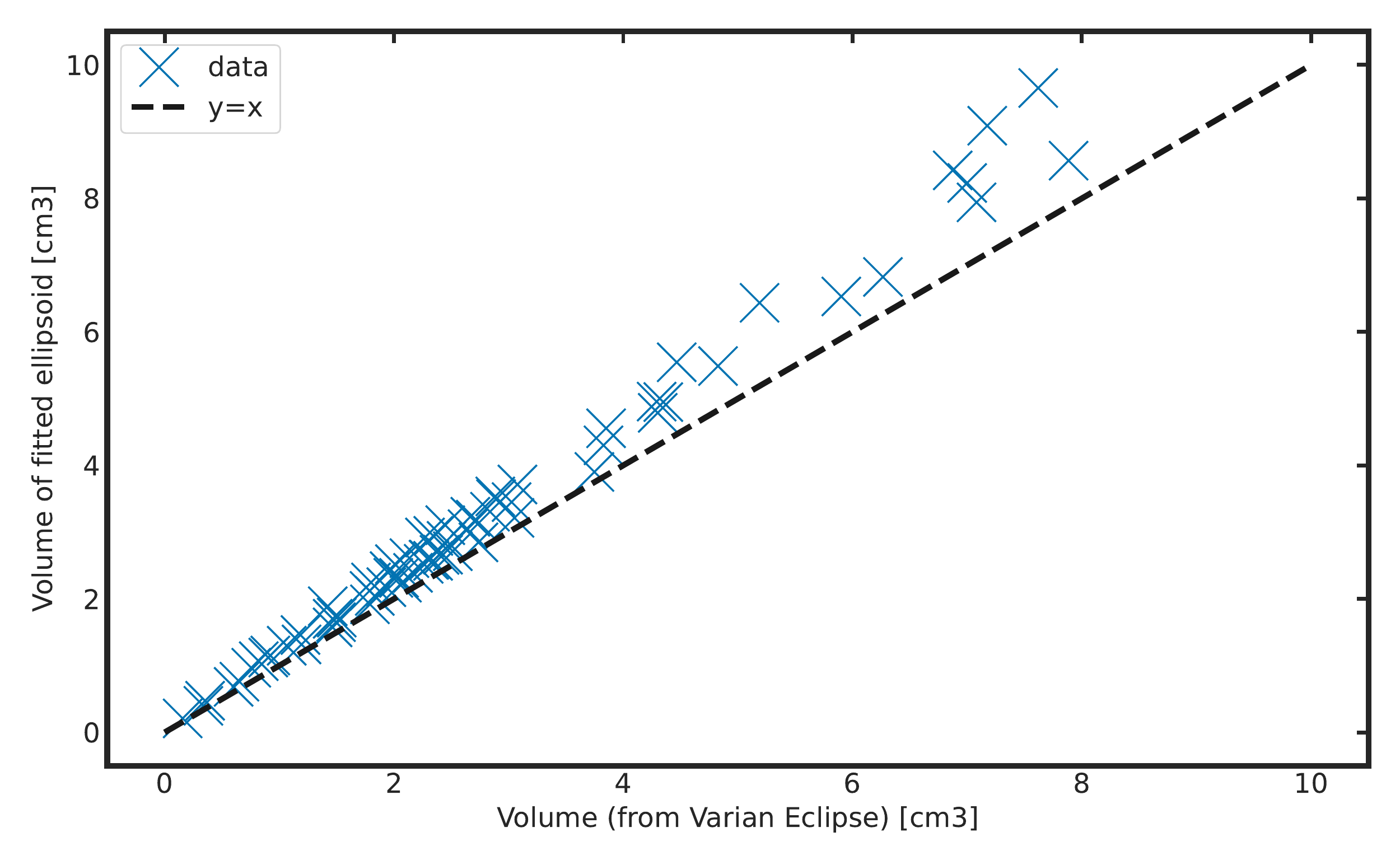}%
}
\caption{\textbf{Volumes of the fitted ellipsoids $V_{ellips}$ are plotted against the actual tumor volumes ($V_{ecl}$ or $V_{est}$.). The dashed line shows $y=x$ axis.}
A: human NSCLC; B: canine meningioma}
\label{fig:vols_ellips_vs_vol_act}
\end{figure}

An illustration of a resulting ellipsoid from fitting to the GTV contour data for a single case is shown in the appendix in Figure~\ref{fig:illu_ellips_3d}.
In Figure~\ref{fig:vols_ellips_vs_vol_act}, volumes of the fitted ellipsoids $V_{ellips}$ are plotted against the actual tumor volumes. By 'actual' tumor volumes for the canine meningioma (Figure~\ref{fig:vols_ellips_vs_vol_act}b) we are refering to volumes of the contoured GTV structures as calculated by Varian Eclipse $V_{ecl}$. For the human NSCLC (Figure~\ref{fig:vols_ellips_vs_vol_act}a) data the volumes $V_{est}$ were estimated as described in methods \& materials.

\begin{figure}[!h]
  \centering
\subfloat[\label{dia_ellips_vs_dist_max:a}]{%
\includegraphics[height=0.33\textheight]{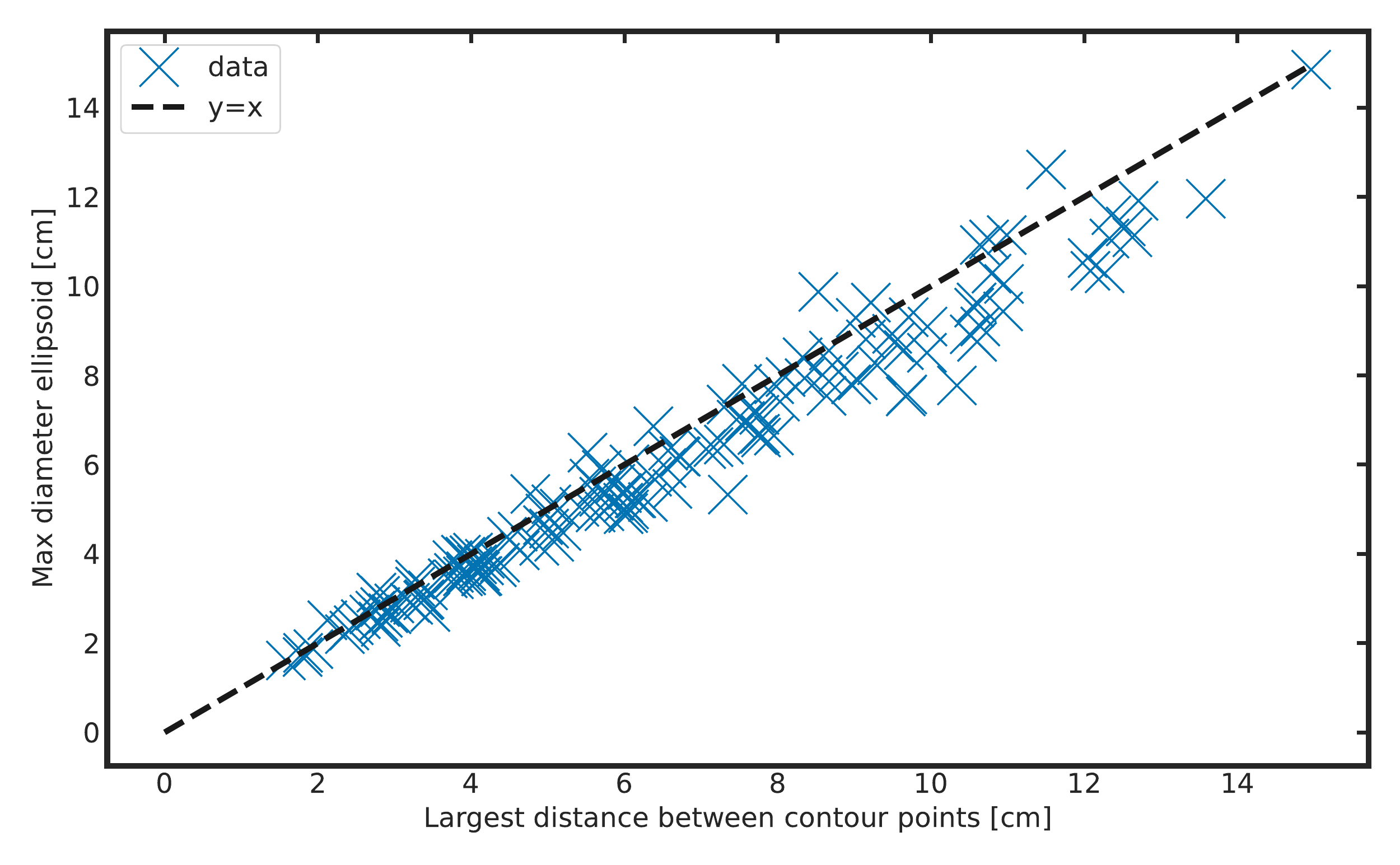}%
}\hfill
\subfloat[\label{dia_ellips_vs_dist_max:b}]{%
\includegraphics[height=0.33\textheight]{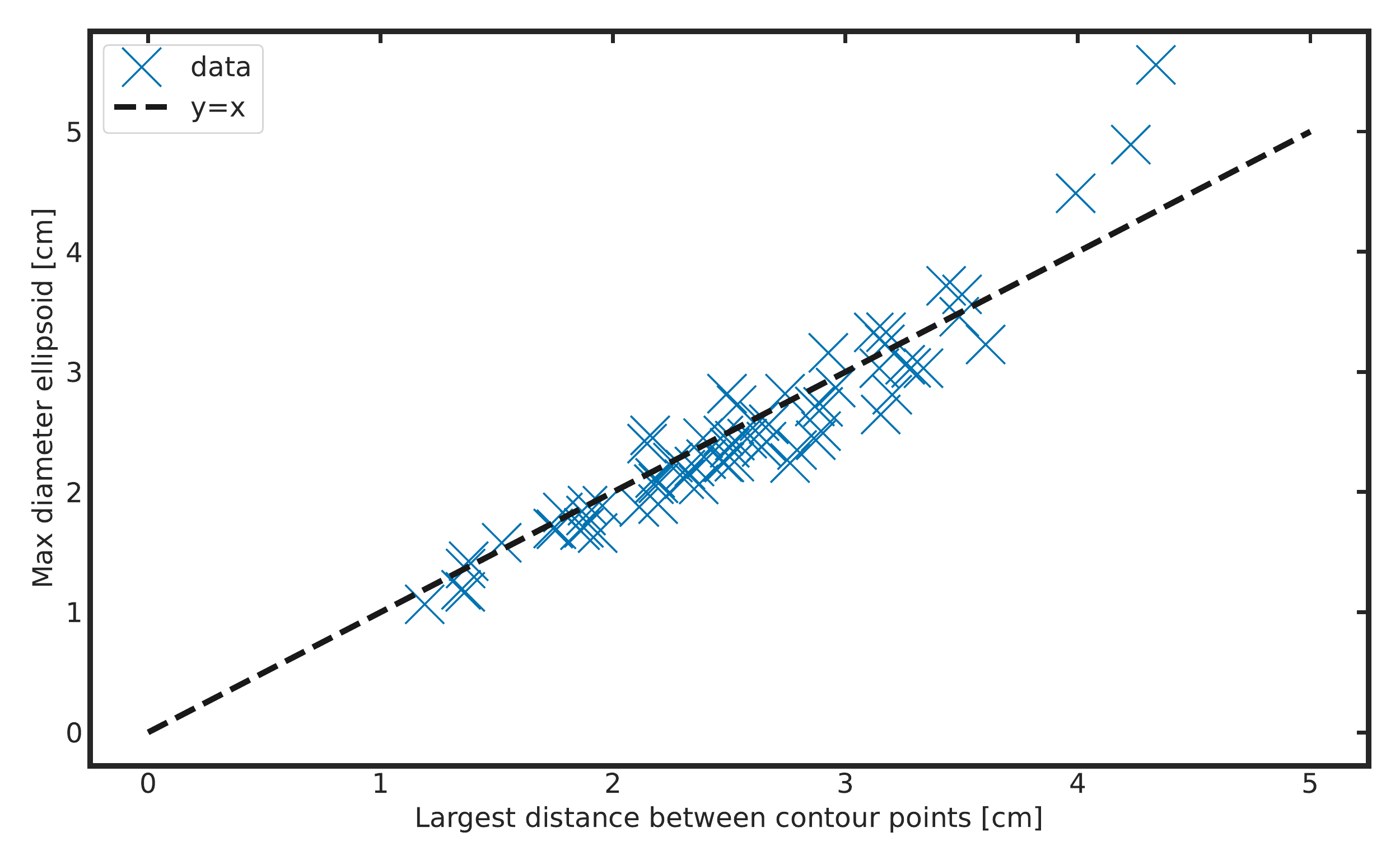}%
}
\caption{\textbf{The largest diameters of the fitted ellipsoids $dia_{max}$  are plotted against the largest distances between contour points $dist_{max}$ of the GTVs. The dashed line shows $y=x$ axis.}
A: human NSCLC; B: canine meningioma}
\label{fig:dia_ellips_vs_dist_max}
\end{figure}

 In Figure~\ref{fig:dia_ellips_vs_dist_max}, the largest diameters of the fitted ellipsoids $dia_{max}$  are plotted against the largest distances between contour points $dist_{max}$ of the GTVs. As can be observed for both the volumes and the largest tumor diameters, the values which stem from the actual tumor structures match quite well to the values of the fitted ellipsoid.
 
\begin{figure}[!h]
 \centering
\subfloat[\label{vol_vs_dmax:a}]{%
\includegraphics[width=0.49\textwidth]{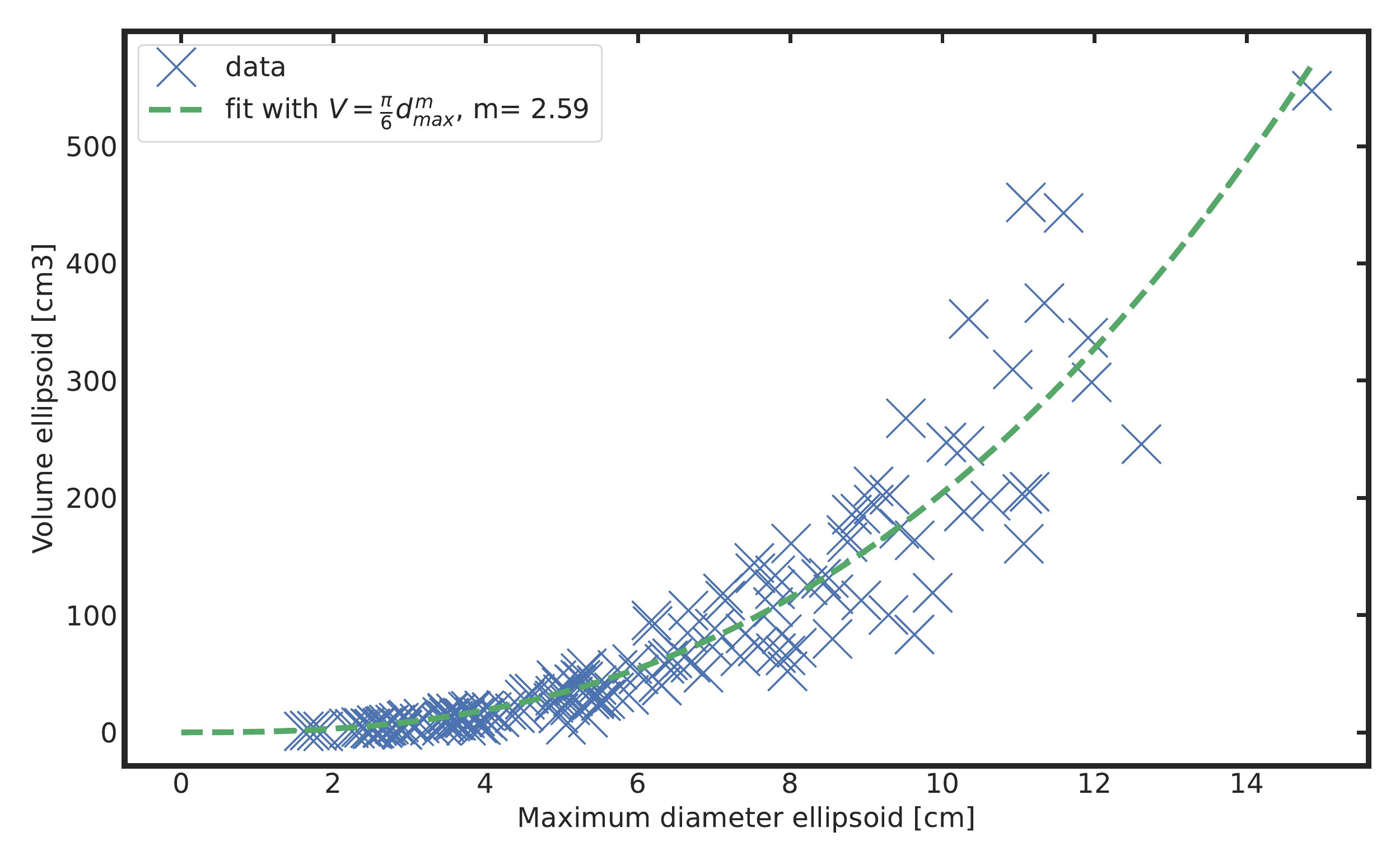}%
}\hfill
\subfloat[\label{vol_vs_dmax:b}]{%
\includegraphics[width=0.49\textwidth]{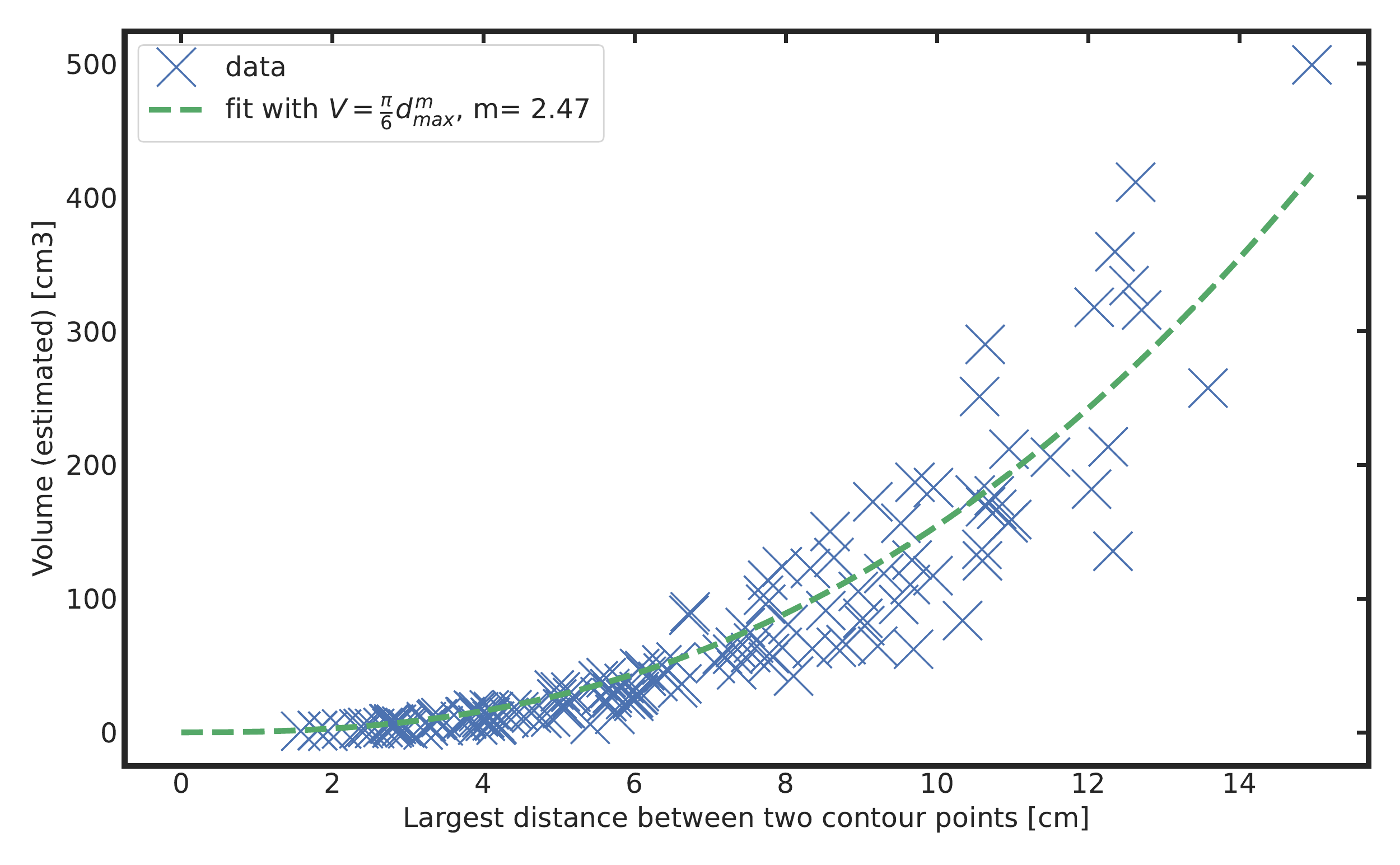}%
}\hfill
\subfloat[\label{vol_vs_dmax:c}]{%
\includegraphics[width=0.49\textwidth]{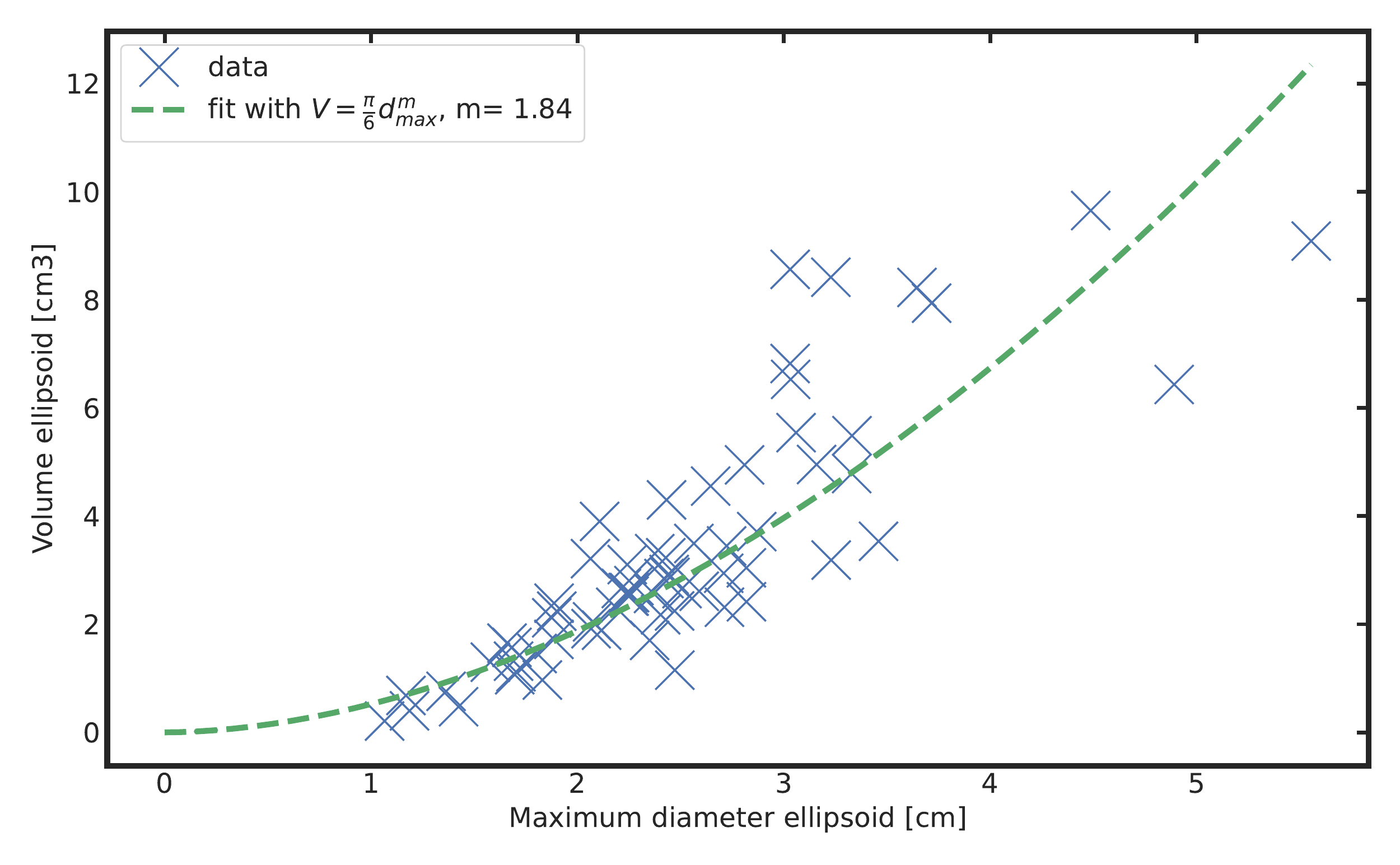}%
}\hfill
\subfloat[\label{vol_vs_dmax:d}]{%
\includegraphics[width=0.49\textwidth]{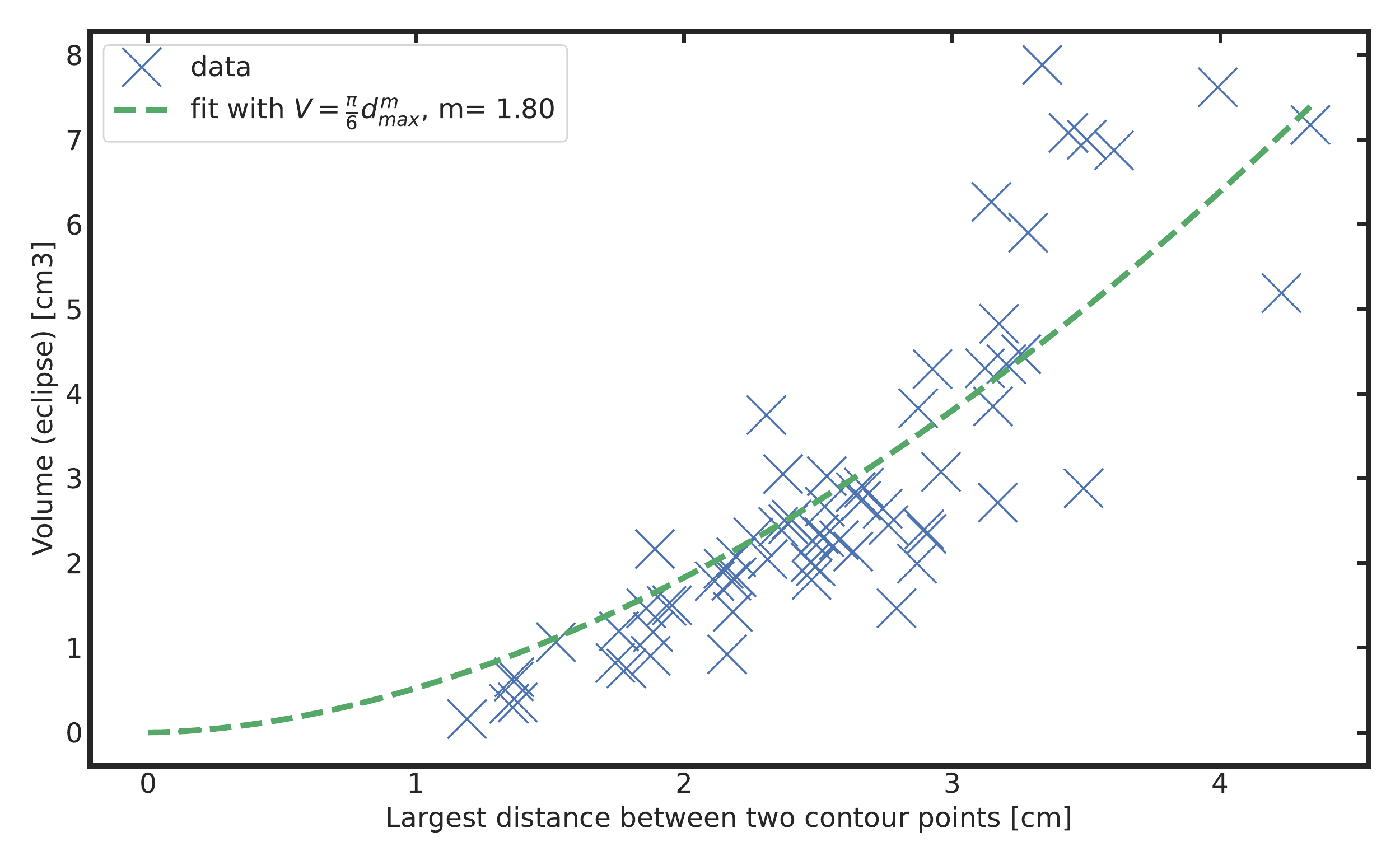}%
}
\caption{\textbf{The volumes are plotted against the maximal diameters. Also the fit with Eq.~\ref{eq:elips_vol} with $\mu=1$ cm is shown.}
A: human NSCLC ellipsoids; B: human NSCLC GTVs; C: canine meningioma ellipsoids; D: canine meningioma GTVs }
\label{fig:vol_vs_dmax}
\end{figure}

In Figure~\ref{fig:vol_vs_dmax}, the volumes of the ellipsoids $V_{ellips}$ and of the actual GTVs ($V_{ecl}$  or $V_{est}$) are plotted against the $dia_{max}$ and $dist_{max}$  respectively. The relation from Eq.~\ref{eq:elips_vol} was fitted to the data yielding the parameter $m$.  
\clearpage
\newgeometry{right=1cm,left=1cm}

\clearpage
\restoregeometry

\begin{figure}[h!]
  \centering
\subfloat[\label{d_mid_d_min_vs_d_max:a}]{%
\includegraphics[height=0.33\textheight]{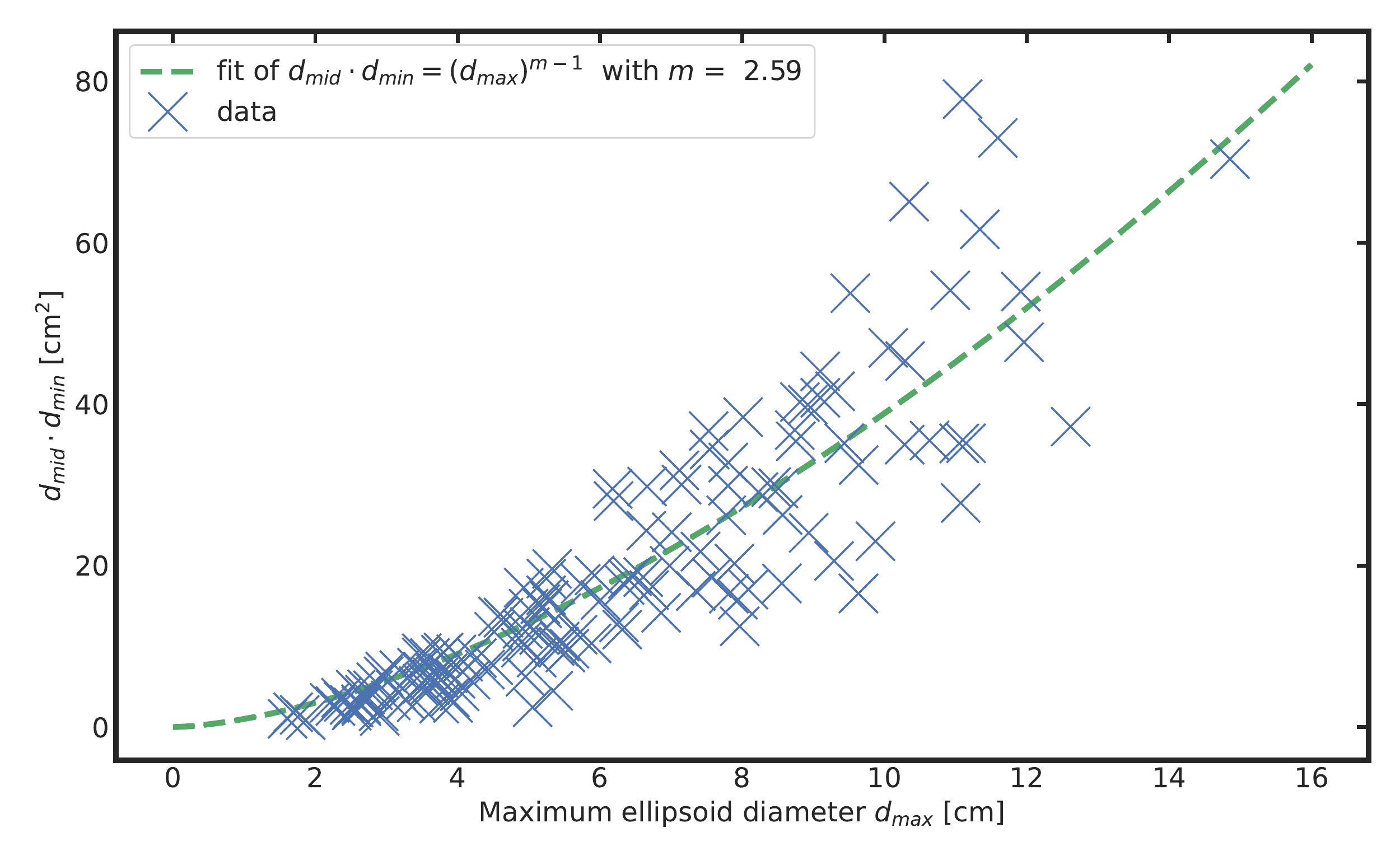}%
}\hfill
\subfloat[\label{d_mid_d_min_vs_d_max:b}]{%
\includegraphics[height=0.33\textheight]{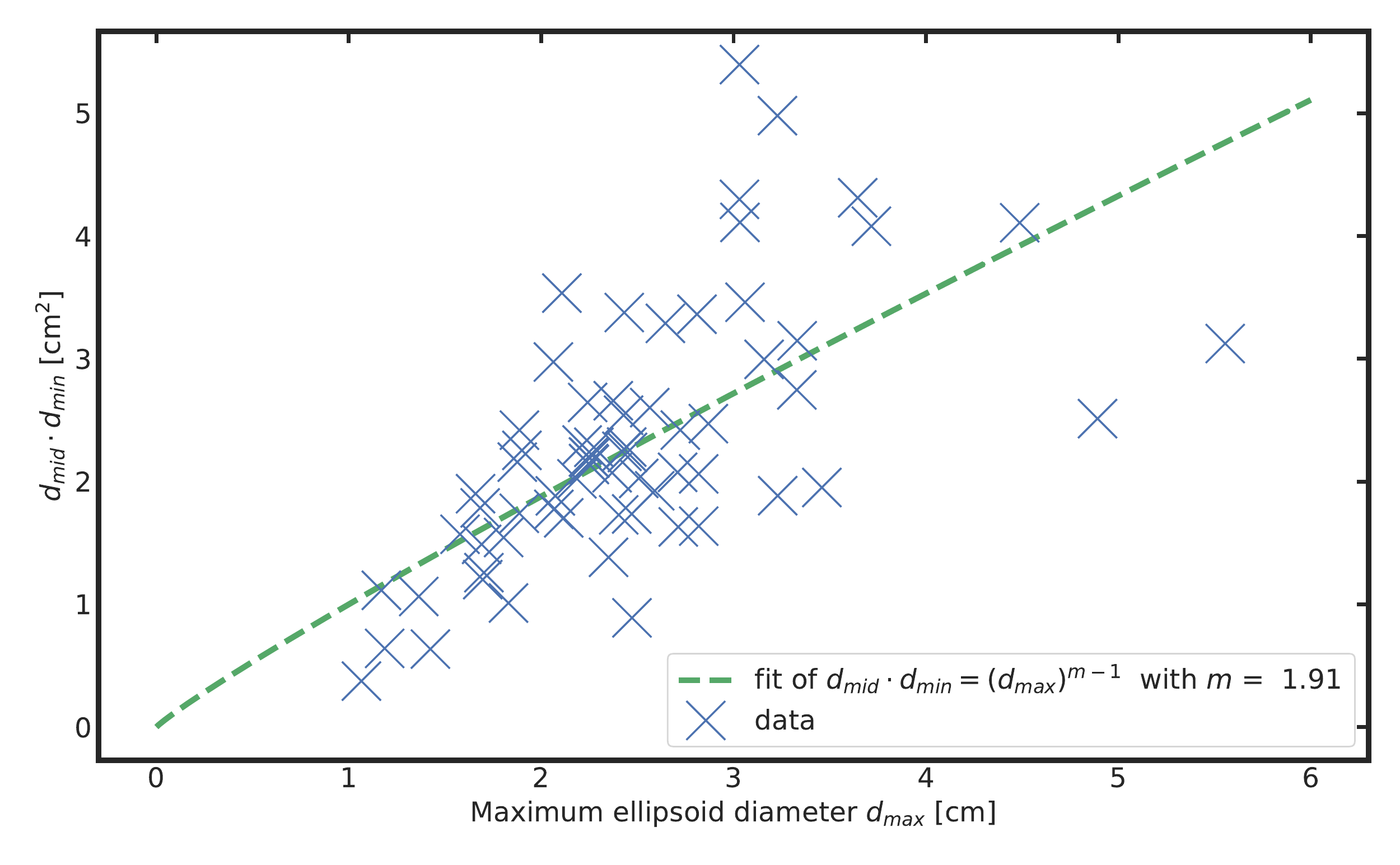}%
}
\caption{\textbf{The squares of the geometrical mean of the smallest and second largest ellipsoid diameter are plotted against the maximal diameters of the ellipsoid. The fit with Eq~\ref{eq:d_max_v_dgeom} with $\mu=1$ cm is shown as the dashed line.}
A: human NSCLC ellipsoids; B: canine meningioma ellipsoids }
\label{fig:d_mid_d_min_vs_d_max}
\end{figure}

Figure~\ref{fig:d_mid_d_min_vs_d_max} shows the square of the geometric mean of the two minor diameters of the ellipsoids plotted against the corresponding maximal diameter. From Eq.~\ref{eq:elips_vol}, the relation between the geometric mean of the two minor ellipsoid axes and the maximal ellipsoid axis
\begin{equation}
d_{mid}\cdot d_{min} = \mu^{3-m}(d_{max})^{m-1}
\label{eq:d_max_v_dgeom}
\end{equation}
can be obtained and fitted to the data. The fits yield a parameter $m$, which is very similar to the $m$ parameters obtained earlier from fitting Eq.~\ref{eq:elips_vol}. 

\begin{figure}[!h]
 \centering
\subfloat[\label{weibull_dmax:a}]{%
\includegraphics[width=0.49\textwidth]{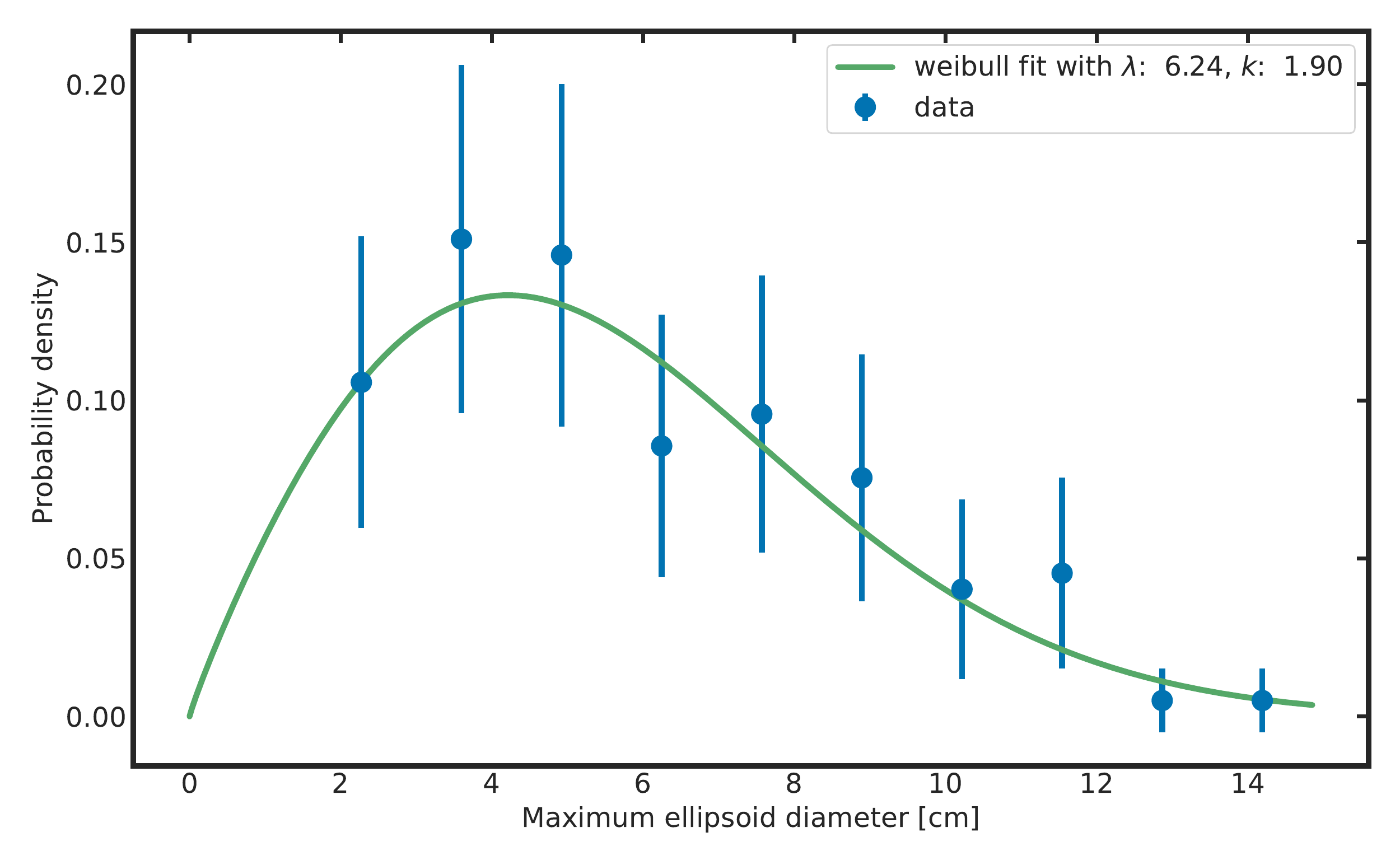}%
}\hfill
\subfloat[\label{weibull_dmax:b}]{%
\includegraphics[width=0.49\textwidth]{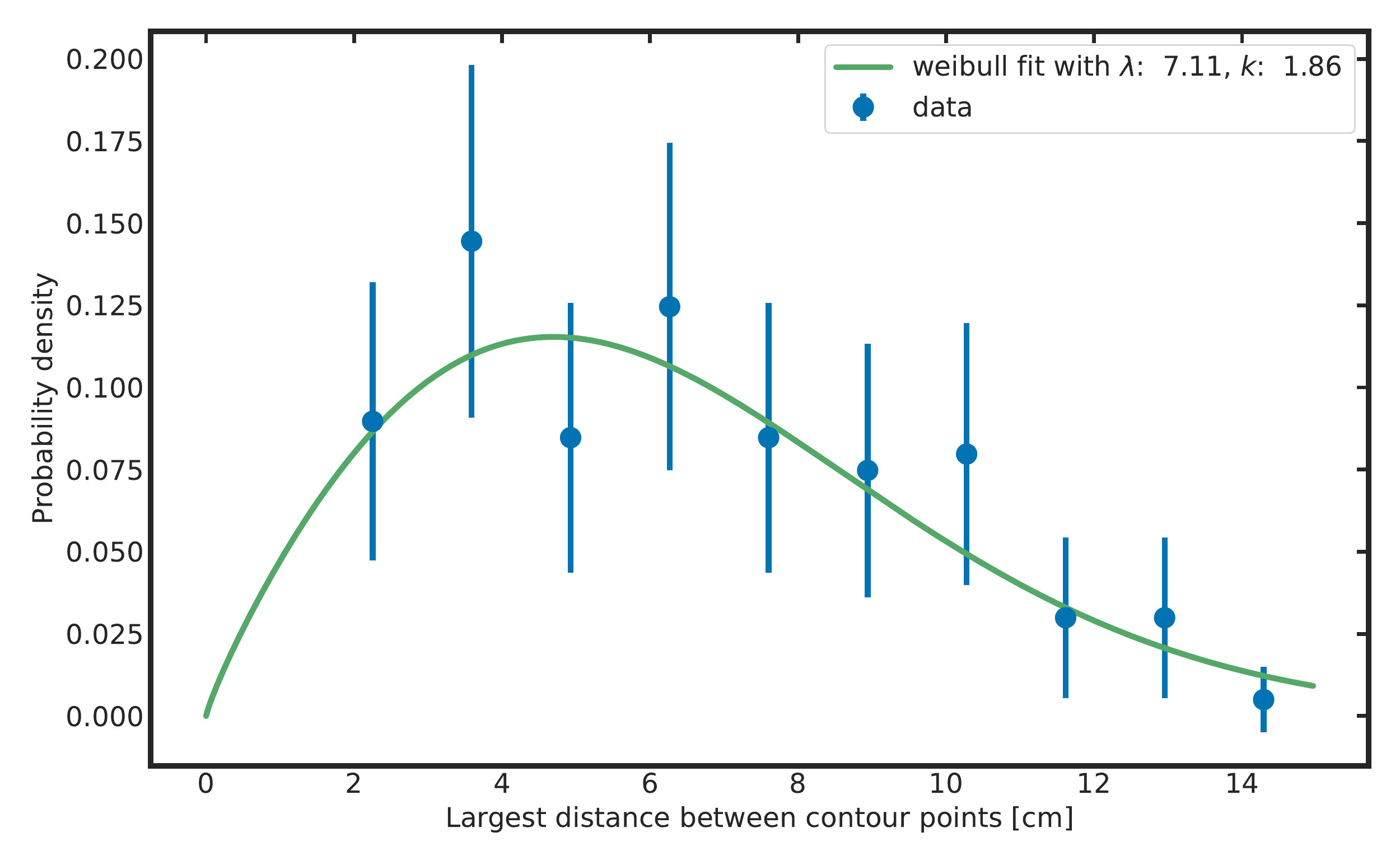}%
}\hfill
\subfloat[\label{weibull_dmax:c}]{%
\includegraphics[width=0.49\textwidth]{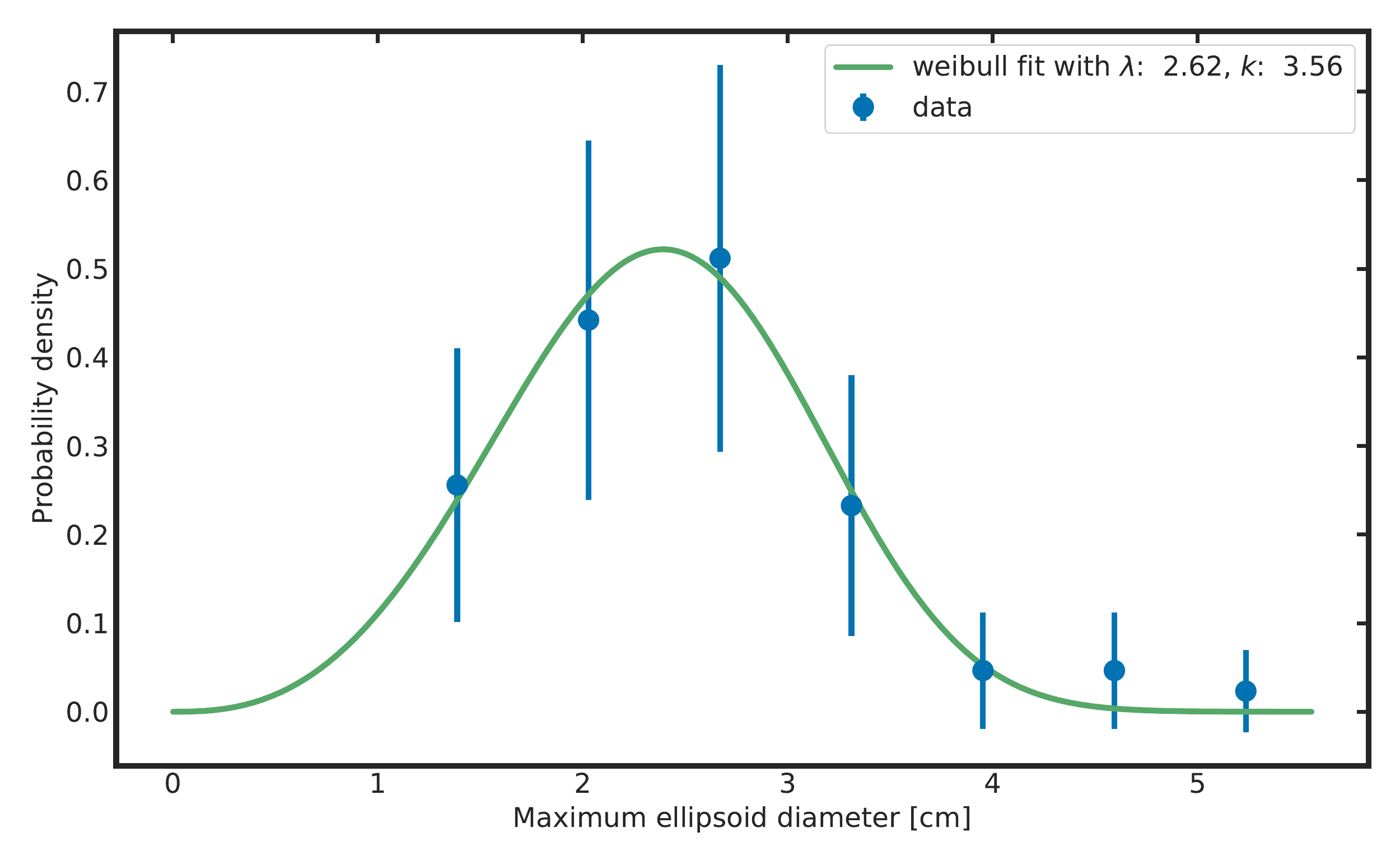}%
}\hfill
\subfloat[\label{weibull_dmax:d}]{%
\includegraphics[width=0.49\textwidth]{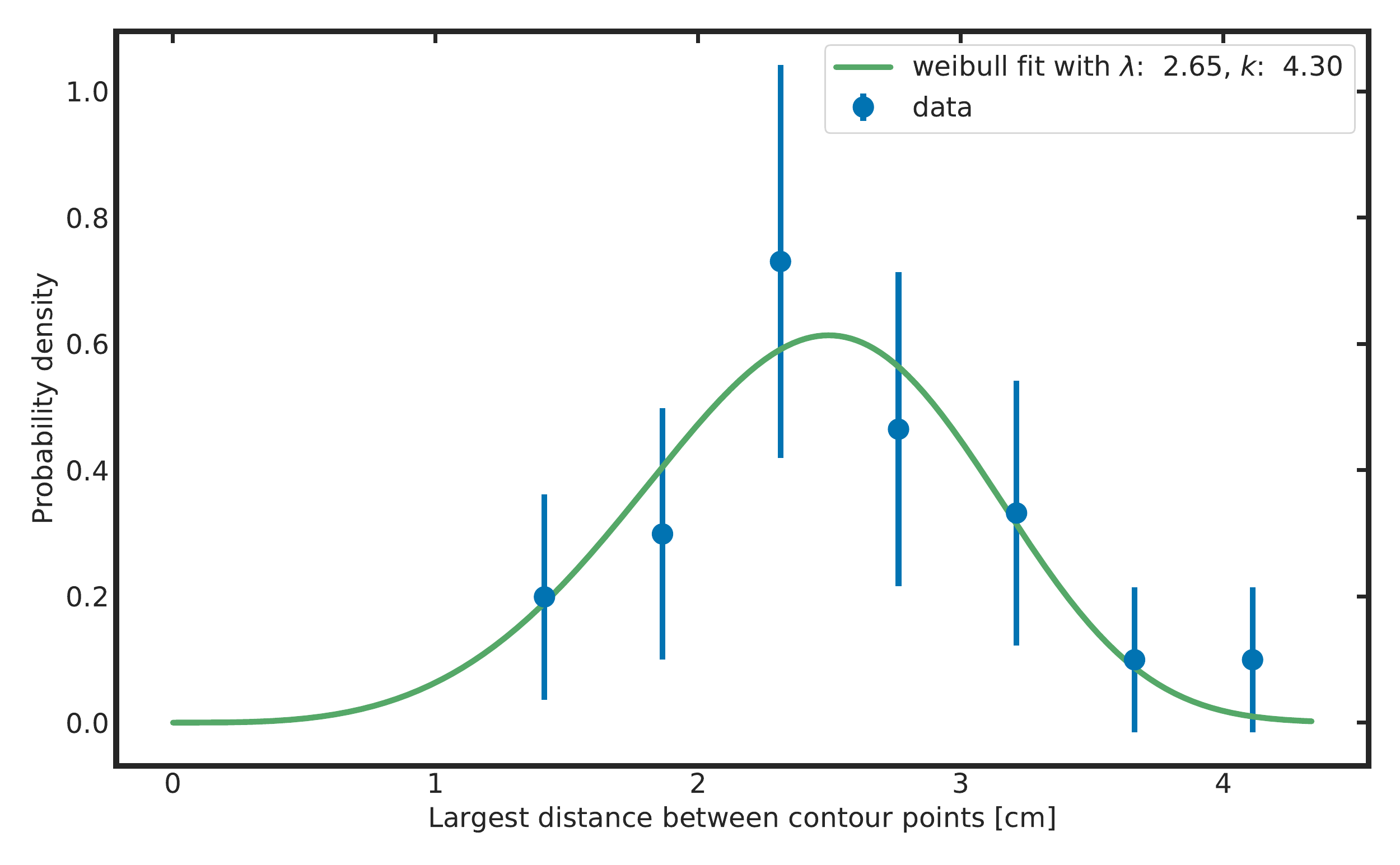}%
}
\caption{\textbf{The histogram of the tumor diameters with the Weibull fits are shown. The errorbars show $2\sigma$ calculated with Eq.~\ref{eq:histogram_err}}
A: human NSCLC ellipsoids; B: human NSCLC GTVs; C: canine meningioma ellipsoids; D: canine meningioma GTVs }
\label{fig:weibull_dmax}
\end{figure}

Figure~\ref{fig:weibull_dmax} shows the normalized histograms of $dia_{max}$ and $dist_{max}$ respectively with the corresponding Weibull fits. The errorbars represent $2\sigma$ (as in Eq.~\ref{eq:histogram_err})

\clearpage
\newgeometry{right=1cm,left=1cm}

\clearpage
\restoregeometry

\begin{figure}[h]
  \centering
\subfloat[\label{weibull_vol_nsclc:a}]{%
\includegraphics[height=0.33\textheight]{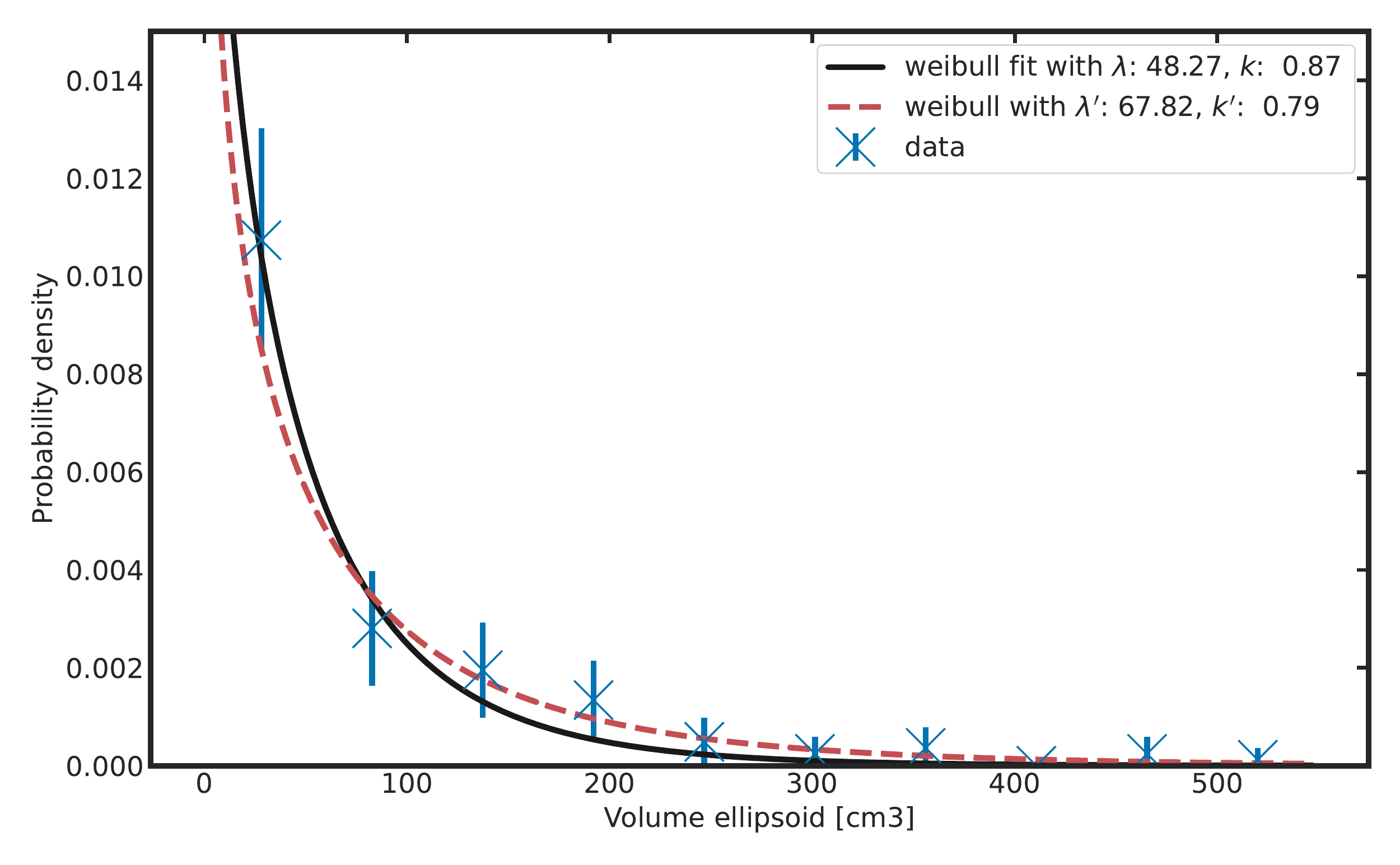}%
}\hfill
\subfloat[\label{weibull_vol_nsclc:b}]{%
\includegraphics[height=0.33\textheight]{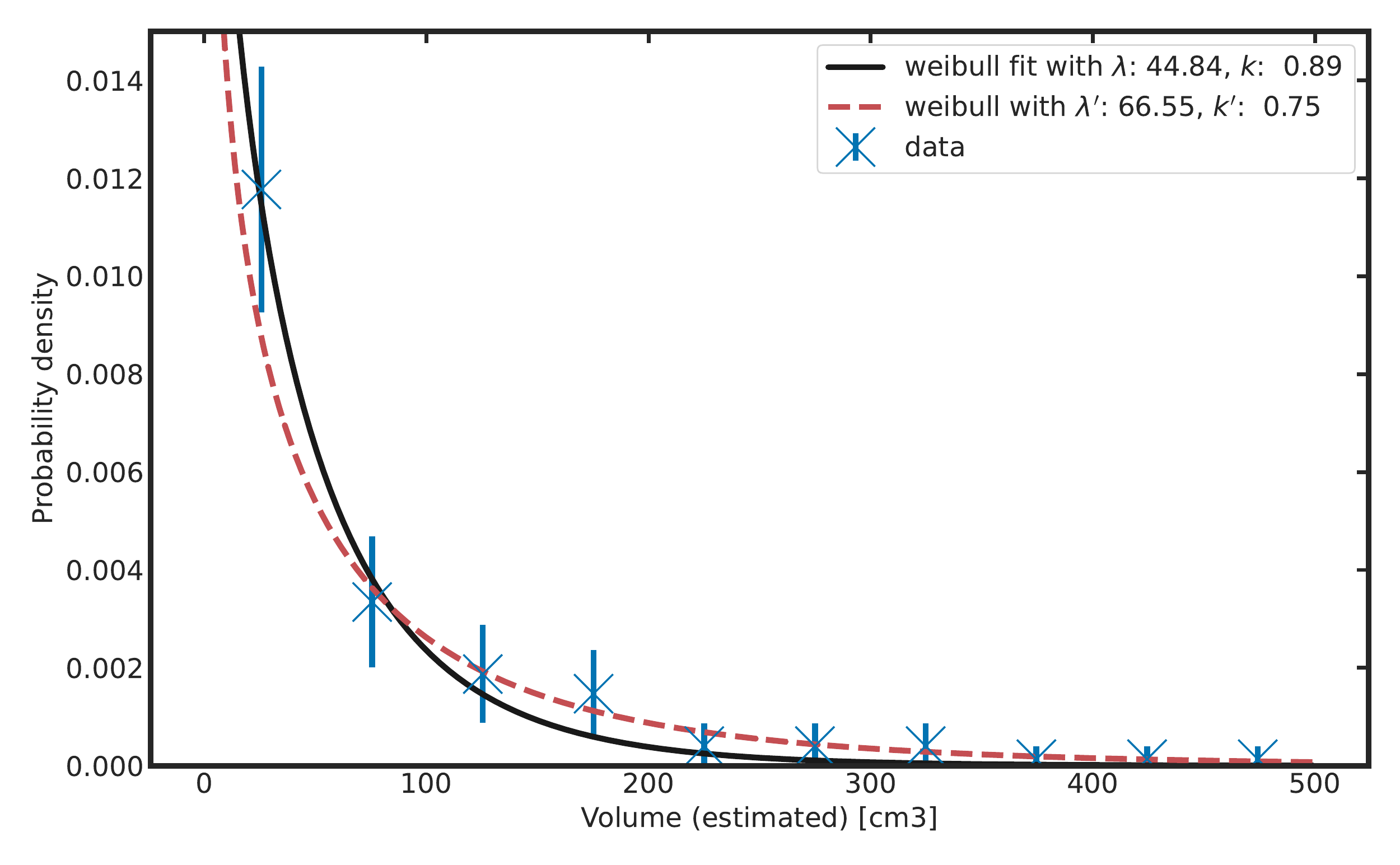}%
}
\caption{\textbf{The histograms of the human NSCLC tumor volumes are shown. The full line is the Weibull fit. The dashed line is the Weibull distribution with calculated parameters $k_V$ and $\lambda_V$. The errorbars show $2\sigma$ calculated with Eq.~\ref{eq:histogram_err} }
A: ellipsoids; B: GTVs; }
\label{fig:weibull_vol_nsclc}
\end{figure}

\begin{figure}[h!]
  \centering
\subfloat[\label{weibull_vol_meng:a}]{%
\includegraphics[height=0.33\textheight]{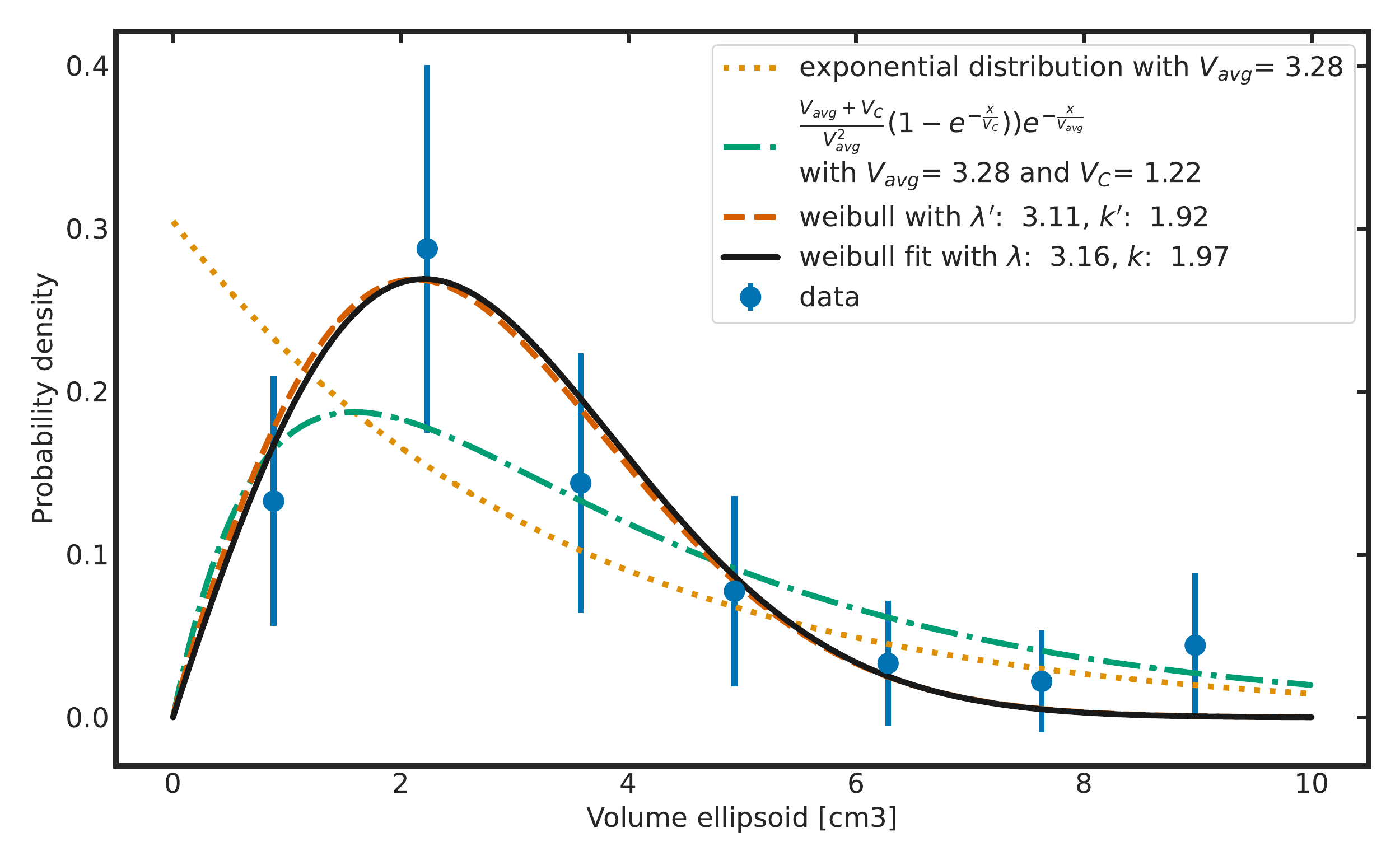}%
}\hfill
\subfloat[\label{weibull_vol_meng:b}]{%
\includegraphics[height=0.33\textheight]{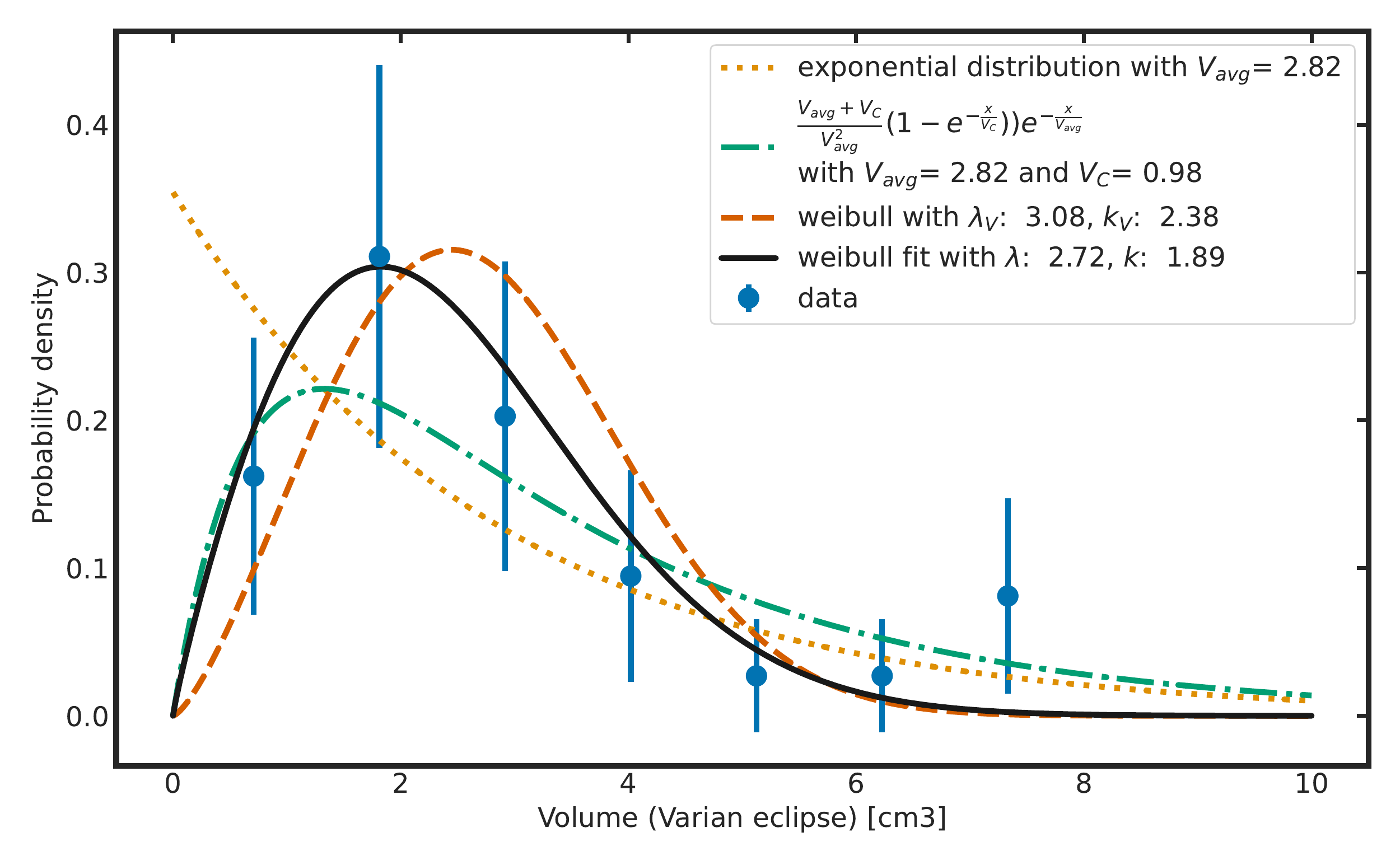}%
}
\caption{\textbf{The histograms of the canine meningioma tumor volumes are shown. The full line is the Weibull fit. The dotted line is the exponential distribution, the dash-dotted line is the exponential distribution with an detection limitation. The dashed line is the Weibull distribution with calculated parameters $k_V$ and $\lambda_V$. The errorbars show $2\sigma$ calculated with Eq.~\ref{eq:histogram_err}. }
A: ellipsoids; B: GTVs; }
\label{fig:weibull_vol_meng}
\end{figure}

Figures~\ref{fig:weibull_vol_nsclc} and \ref{fig:weibull_vol_meng} show the normalized histograms of $V_{ecl}$ or $V_{est}$ and $V_{ellips}$. 
Also plotted are a Weibull distribution fit and a Weibull distribution with parameters $\lambda_V$ and $k_V$ calculated from parameters obtained from Weibull fits to the distributions of $dia_{max}$ and $dist_{max}$.

Table~\ref{table:res_overview} shows an overview of the resulting fitting parameters.

\begin{table}[h!]
\begin{tabular}{@{}lccc|cc@{}}
\toprule
\textbf{}                                                                                            & \multicolumn{1}{l}{}                 & \multicolumn{2}{c|}{\textbf{NSCLC}}                & \multicolumn{2}{c}{\textbf{(canine) meningioma}} \\
\textbf{}                                                                                            & \multicolumn{1}{l}{}                 & GTV               & \multicolumn{1}{l|}{ellipsoid} & GTV                     & ellipsoid              \\ \midrule
\multirow{2}{*}{Weibull fits of diameter}                                                            & $k$                                  & $1.86 \pm 0.21$   & $1.90 \pm 0.17$                & $4.30 \pm 0.53$         & $3.56 \pm 0.17$        \\
                                                                                                     & $\lambda$                            & $7.11 \pm 0.51$   & $6.24 \pm 0.33$                & $2.66 \pm 0.09$         & $2.62 \pm 0.04$        \\ \midrule
\multirow{2}{*}{Weibull fits of volume}                                                              & $k_V$                                  & $0.89 \pm 0.06$   & $0.88 \pm 0.06$                & $1.89 \pm 0.21$         & $1.97 \pm 0.21$        \\
                                                                                                     & $\lambda_V$                            & $44.84 \pm 5.40$  & $48.27 \pm 7.15$               & $2.72 \pm 0.21$         & $3.16 \pm 0.23$        \\ \midrule
\multirow{2}{*}{\begin{tabular}[c]{@{}l@{}}Calculated Weibull \\ parameters for volume\end{tabular}} & $k_V$                                & $0.75 \pm 0.09$   & $0.79 \pm 0.07$                & $2.38 \pm 0.30$         & $1.92 \pm 0.1$         \\
                                                                                                     & $\lambda_V$                          & $66.55 \pm 11.86$ & $67.82 \pm 8.95$               & $3.08 \pm 0.21$         & $3.11 \pm 0.15$        \\ \midrule
Parameter $m$ fit                                                                                    & $V=\frac{\pi}{6}(d_{max})^m$         & $2.47 \pm 0.01$   & $2.59 \pm 0.01$                & $1.81 \pm 0.03$         & $1.85\pm 0.03$         \\
                                                                                                     & $d_{min}\cdot d_{mid}=d_{max}^{m-1}$ & -                 & $2.59 \pm 0.01$                & -                       & $1.91\pm0.04$          \\ \midrule
                                                                                                     & $\overline{V}$ {[}cm$^3${]}          & $70.29 \pm 0.59$  & $78.89 \pm 0.66$               & $2.82 \pm 0.03$         & $3.29 \pm 0.03$        \\
\begin{tabular}[c]{@{}l@{}}Exp. Distr. with \\ detection limitation\end{tabular}                     & $V_C$ {[}cm$^3${]}                   & -                 & -                              & $0.98 \pm 0.59$         & $1.22 \pm 0.81$       
\end{tabular}
\caption{Overview of the obtained parameters}
\label{table:res_overview}
\end{table}

\subsection{Tumor growth simulations}

\begin{figure}[!h]
 \centering
\subfloat[\label{weibull_vol_nsclc_sim_seer:a}]{%
\includegraphics[height=0.33\textheight]{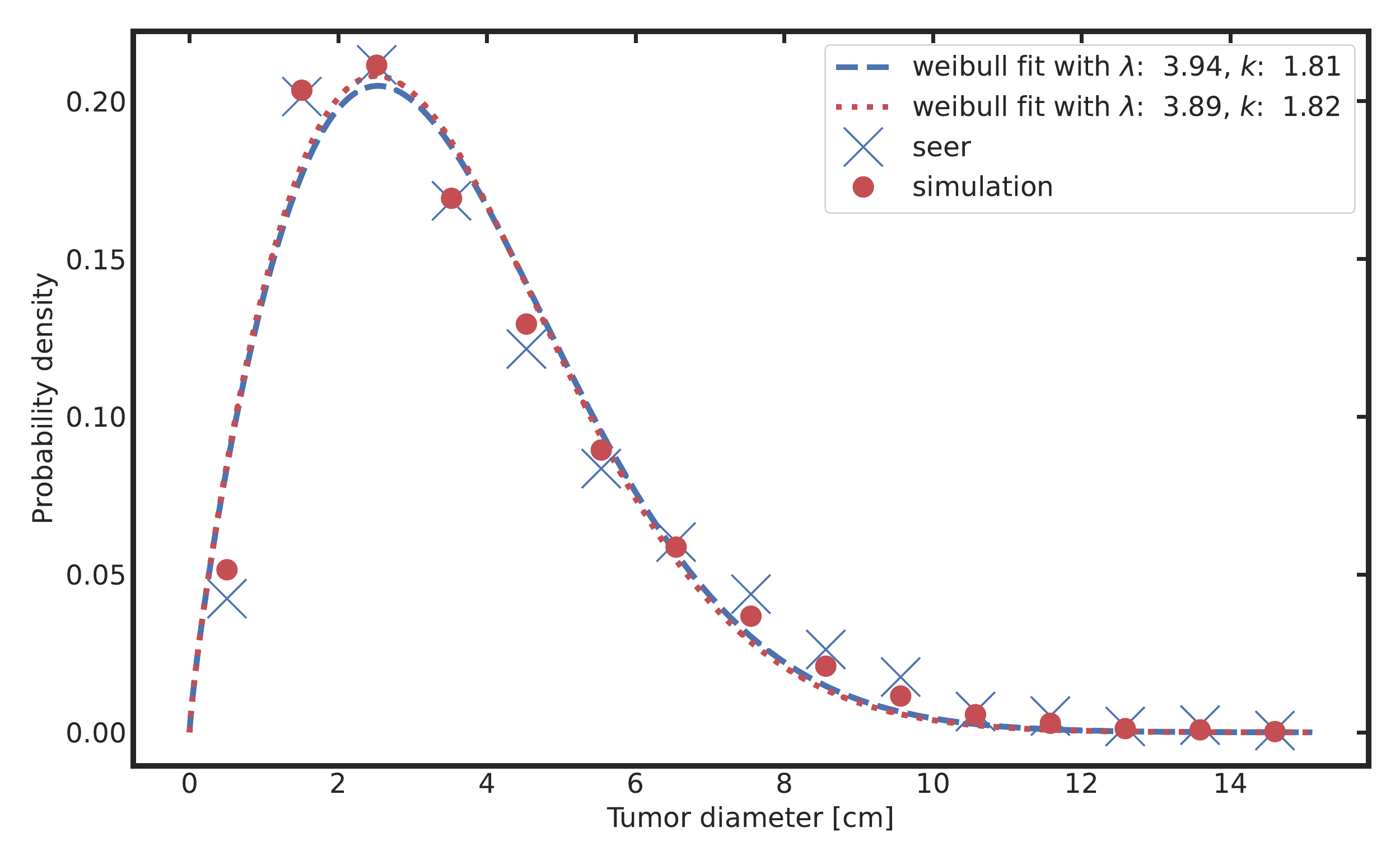}%
}\hfill
\subfloat[\label{weibull_vol_nsclc_sim_seer:b}]{%
\includegraphics[height=0.33\textheight]{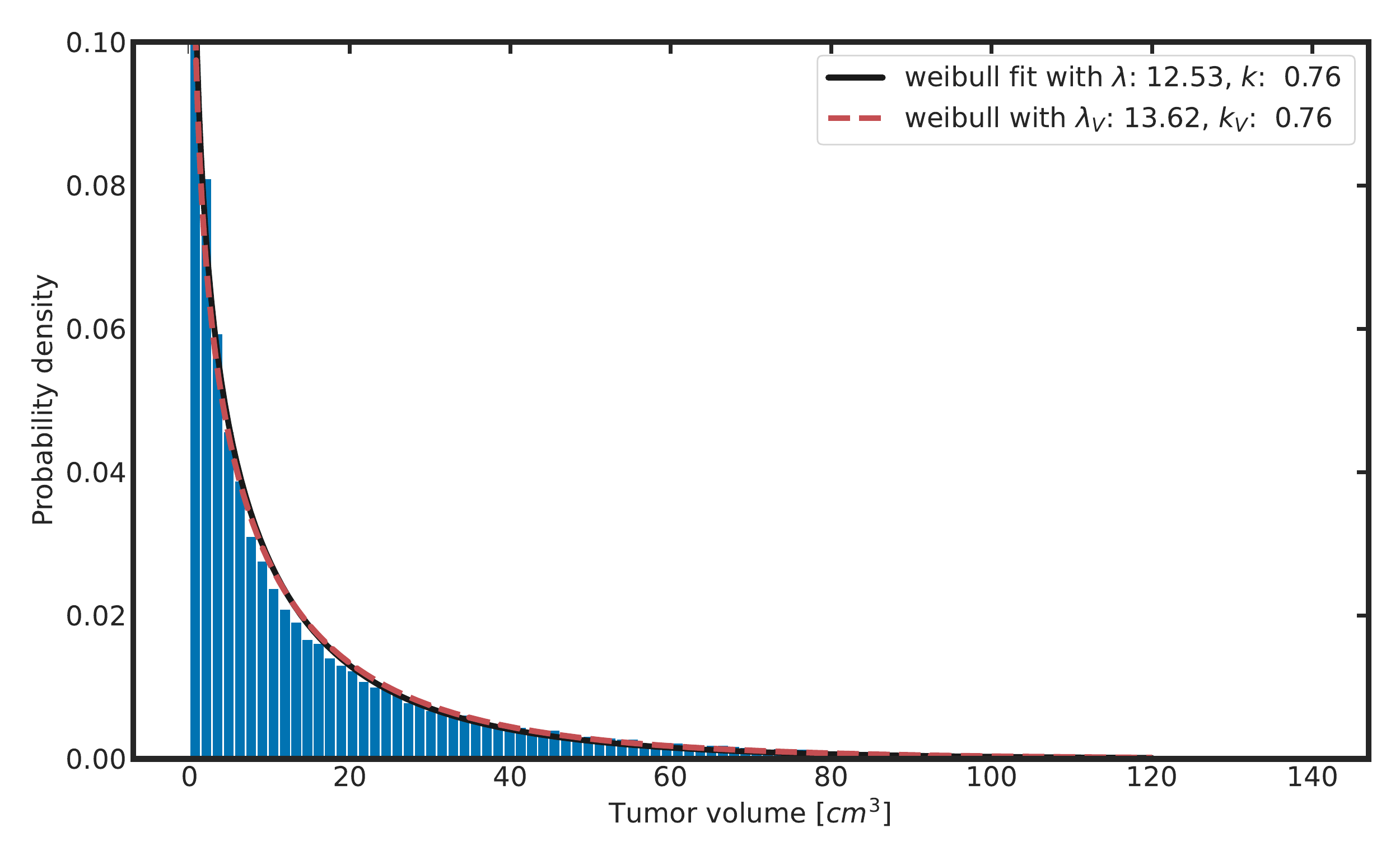}%
}
\caption{
\textbf{A: Histograms of NSCLC tumors diameters from simulation are compared with SEER data}, $\times$ and $\bullet$ show the histogram data points from SEER and simulation respectively,  dashed and dotted lines show the Weibull fit to SEER and simulation data respectively; \textbf{B: Histogram of tumor volumes from simulation}, solid line shows the Weibull fit, dashed line shows the Weibull distribution with calculated parameters $k_V$ and $\lambda_V$ }
\label{fig:weibull_vol_nsclc_sim_seer}
\end{figure}

In Figure~\ref{fig:weibull_vol_nsclc_sim_seer}A the NSCLC tumor diameter distribution detected at an age of 70 years, resulting from the Monte Carlo simulation is compared to the diameter distribution from SEER. Also the Weibull fits to the corresponding data are plotted. Figure~\ref{fig:weibull_vol_nsclc_sim_seer}B shows the NSCLC tumor volume distribution detected at age of 70 years, resulting from the Monte Carlo simulation. Also plotted is the Weibull fit, as well as the Weibull distribution with computed $k_V$ and $\lambda_V$ (Eq.~\ref{eq:k_vol} and \ref{eq:lambda_vol}). As explained in the corresponding methods \& material section, the fitting and simulation procedures were done both using a constant $m$ for all tumors, as well as by assuming $m$ to be normally distributed. Both variants resulted in very similar growth rate and detection limitation parameters $r=0.36\pm0.04$ year$^{-1}$ and $V_C=0.6\pm0.15$ cm$^3$. The latter was done in an attempt make the simulation more realistic. However, both variants yielded results which were very close to each other.

\clearpage

\section{Discussion}

In previous research the distribution of tumor size and volume in patient populations remained widely unexplored. In this work, SEER data, as well as three dimensional CT data from patients undergoing radiation therapy were used to explore and model distributions of tumor diameters and volumes.

It was shown for multiple types of cancer, that the Weibull distribution can adequately represent how the largest diameters of tumors are distributed within the patient population. 
Under the assumption of an ellipsoidal tumor shape and that the maximal tumor diameters are Weibull distributed, it was mathematically shown that the tumors volumes are Weibull distributed as well. Therefore, the parameters of the derived volume Weibull distribution can be calculated from parameters obtained from Weibull fits to the largest diameters. In case where $k_V=\gamma=1$, the tumor volumes are exponentially distributed. \\
\\
As can be observed in Figure~\ref{fig:dia_ellips_vs_dist_max} , for both the human NSCLC and canine meningioma the $dia_{max}$ of the fitted ellipsoids match the $d_{max}$ of the GTVs quite well, all data points being close to the $y=x$ line. 
The same observation holds true for the volumes of the ellipsoids and GTVs, as can be seen in Figure~ \ref{fig:vols_ellips_vs_vol_act}.
It can be observed that the ellipsoid fits slightly overestimate the actual GTV volumes. For the NSCLCs the maximal diameters of the ellipsoids are slightly smaller than the largest distances between contour points of the GTVs. 
Further, fitting the volume diameter relation in Figure \ref{fig:vol_vs_dmax} yields the parameter $m$ which describes the geometric shape of the tumors under the assumption of equation \eqref{eq:elips_vol}. It is very similar for both fits to the GTV and to ellipsoid data. 
\\
\\
As already seen for the SEER data, in Fig.~\ref{fig:weibull_dmax} it can be observed that the Weibull distributions are also a good fit for the largest diameters obtained from analysed 3D data. The Weibull distribution with $k_V$ and $\lambda_V$ calculated from parameters obtained from the Weibull fits of the diameters, and with the previously determined $m$ parameter matches the observed tumor volume distributions, as shown in Fig.~\ref{fig:weibull_vol_nsclc} and \ref{fig:weibull_vol_meng}. The observed volume distribution of the canine meningioma shows a reduced probability of detection for very small volumes. In contrast to that, for NSCLC volumes this effect is not visible. We attribute this to a detection limitation for small volumes. In this context we believe it might be reasonable to conjecture that the detection limitation in canine meningioma is a much more visible effect as in contrast to humans, as there are no preventive screenings, and symptoms are noticed much later. \\
\\
The analysis of diameters and volumes of the fitted ellipsoids was constrained to cases where the difference of volume of the fitted ellipsoid and volume of the actual GTV was less than 35\%. This criterion was met for 90\% of the NSCLC and for 72\% of the canine meningioma cases. The constraint was necessary, as the method used for fitting the ellipsoids to the contour points does not always produce ellipsoids which resemble the actual tumor structure. 
Not all tumors can be approximated by ellipsoids appropriately. However, for the majority of human NSCLC and canine meningioma data sets, investigated in this work, the ellipsoid can adequately approximate the tumor shape.\\
\\

Three dimensional tumor contours contain exact information about tumor geometry.
However, generally the analysis of three dimensional data bears certain difficulties. To begin with, there are not many suitable contoured GTV datasets available. Usually, contoured three dimensional data stem only from patients undergoing radiation therapy. In most cases, human patients who are irradiated have had previous surgery, the contoured GTV in these cases do not represent the true tumor volume and geometry. Hence, such datasets are unsuitable for the analysis in the scope of this work. Further, data from RT patients might not be representative for the whole patient population.  In order to obtain sufficiently large number of data-points automatic processing is necessary. Sometimes multiple disjoint structures are contoured within a single GTV object or the contours contain some unindented artefacts. Also for some tumor cases multiple GTV structures are present. Such obstacles hinder the automatic processing of datasets. \\
\\

As shown in Figure~\ref{fig:weibull_vol_nsclc_sim_seer}, the observed NSCLC cancer diameter distribution from SEER can be fairly well simulated using the power law growth model and the population carcinogenic cell incidence model from \citet{Shuryak2009, Shuryak2009_2}. It can be observed that histogram data points of the simulation are very close to the corresponding SEER data points. Also the Weibull fits yield very similar parameters. The histogram of the simulated tumor volumes follows the predicted Weibull distribution.
It is noteworthy, that the presented procedure also allows for estimating parameters for tumor growth models from SEER data. \\
\\
Regarding the assumed limited detectability of small tumour sizes (volumes or diameters), our hypothesis in this work was that the Weibull distributions can implicitly model this phenomenon. Hence, the limited detectability of tumours of small size inherent in a given dataset is implicitly included in the Weibull distribution parameters $\lambda$ and $k$. For the tumor growth simulation, Eq.~\ref{eq:detectmodel} was used as a model for the limited detectability. Although the use of this model for the simulation gives a good agreement with the SEER data, it is merely an assumption, which also requires further validation. Other models might provide different insights and/or potentially offer a better fit to the observed reality.
\\
\\
In future work, more three dimensional datasets should be acquired and analysed. An emphasis should also be put on using and comparing more sophisticated tumor growth models such as \citet{10.1371/journal.pone.0022973}. The effect of detection limitation for small volumes should be investigated more closely as well.

\section*{Acknowledgment}
This work was supported by the Swiss National Science Foundation (SNSF), grant number: 320030-182490; PI: Carla Rohrer Bley

\appendix
\renewcommand{\thesubsection}{\Alph{section}.\arabic{subsection}}
\setcounter{subsection}{0}
\renewcommand\thefigure{\thesection.\arabic{figure}}    

\section{Detailed methods}
\subsection*{Fitting ellipsoids to tumor structures}
For this purpose the procedure presented in \citet{Ying2012} was used. 
In a two dimensional cartesian coordinate system, a conic section described by 
\begin{equation}
ax^2+by^2 + cxy + dx + ey + f = 0
\label{eq:conic}
\end{equation}
For a non degenerate conic section, if $4ab-c^2 >0$ the conic section is an ellipse. In a three dimensional cartesian coordinate system, a quadric surface is given by 
\begin{equation}
ax^2+by^2+cz^2+dxy+eyz+fzx+gy+hy+iz+j = 0
\label{eq:quadric}
\end{equation}
Dependent on the parameters, the structure quadric surface corresponds to a standard-form type such as a ellipsoid, hyperboloid, sphere etc.  \citet{Ying2012} state that if a quadric surface is an ellipsoid, then its intersection with a plane is a real or imaginary ellipse. The equation for a random plane in 3D can be written as
\begin{equation}
z = \alpha x + \beta y + \gamma
\end{equation}
, substituting Eq.~\ref{eq:conic} into Eq.~\ref{eq:quadric} yields 
\begin{equation}
Ax^2+By^2+Cxy+Dx+Ey+F=0
\end{equation}
where 
\begin{align}
A &= a+\alpha^2 c + \alpha f\\
B &= b + \beta^2 c+\beta e\\
C &= 2\alpha\beta c + d + \alpha e  \beta f\\
D &= 2\alpha\gamma c + \gamma f + g + \alpha i\\
E &= 2\beta\gamma c + \gamma e + h + \beta i\\
F &= \gamma^2 c + \gamma i + j
\end{align}
Now we have the constraint that $4AB-C^2>0$, which can be rewritten as
\begin{equation}
4AB-C^2 = \mathbf{a}^T\mathbf{M}\mathbf{a}
\end{equation}
where $\mathbf{a}=[a,b,c,d,e,f,g,h,i,j]^T$ and 
\begin{equation}
\mathbf{M} = 
\begin{bmatrix}
0 & 2 & 2\beta^2 &0 & 2\beta & 0 & 0 & 0 & 0 &0\\
2 & 0 & 2\alpha^2 & 0 & 0  & 2\alpha & 0 & 0& 0 & 0\\
2\beta^2 & 2\alpha^2 & 0 & -2\alpha\beta &0 & 0& 0& 0& 0&0  \\
0 & 0 & -2\alpha\beta & -1 & -\alpha & -\beta & 0 & 0 & 0 & 0 \\
2\beta & 0 & 0 & -\alpha & -\alpha^2 & \alpha\beta  & 0 & 0 & 0 & 0 \\
0 & 2\alpha & 0 & -\beta & \alpha\beta & -\beta^2 & 0 & 0 & 0 & 0 \\
0 & 0 & 0 & 0 & 0 & 0 & 0 & 0 & 0 & 0 \\
0 & 0 & 0 & 0 & 0 & 0 & 0 & 0 & 0 & 0 \\
0 & 0 & 0 & 0 & 0 & 0 & 0 & 0 & 0 & 0 \\
0 & 0 & 0 & 0 & 0 & 0 & 0 & 0 & 0 & 0 \\
\end{bmatrix}
\end{equation}
In order to solve for a potential ellipsoidal solution we have 
\begin{equation}
\min \psi = ||\mathbf{D}\mathbf{a}||^2
\end{equation}
so that
\begin{equation}
\mathbf{a}^T\mathbf{M}\mathbf{a} = 1
\end{equation} where
\begin{equation}
\mathbf{D}=\left[\begin{array}{cccccccccc}
\mathrm{x}_1^2 & \mathrm{y}_1^2 & \mathrm{z}_1^2 & \mathrm{x}_1 \mathrm{y}_1 & \mathrm{y}_1 \mathrm{z}_1 & \mathrm{z}_1 \mathrm{x}_1 & \mathrm{x}_1 & \mathrm{y}_1 & \mathrm{z}_1 & 1 \\
\mathrm{x}_2^2 & \mathrm{y}_2^2 & \mathrm{z}_1^2 & \mathrm{x}_2 \mathrm{y}_2 & \mathrm{y}_2 \mathrm{z}_2 & \mathrm{z}_2 \mathrm{x}_2 & \mathrm{x}_2 & \mathrm{y}_2 & \mathrm{z}_2 & 1 \\
\mathrm{x}_3^2 & \mathrm{y}_3^2 & \mathrm{z}_1^2 & \mathrm{x}_3 \mathrm{y}_3 & \mathrm{y}_3 \mathrm{z}_3 & \mathrm{z}_3 \mathrm{x}_3 & \mathrm{x}_3 & \mathrm{y}_3 & \mathrm{z}_3 & 1 \\
\vdots & \vdots & \vdots & \vdots & \vdots & \vdots & \vdots & \vdots & \vdots & \vdots \\
\mathrm{x}_{\mathrm{N}}^2 & \mathrm{y}_{\mathrm{N}}^2 & \mathrm{z}_1^2 & \mathrm{x}_{\mathrm{N}} \mathrm{y}_{\mathrm{N}} & \mathrm{y}_{\mathrm{N}} \mathrm{z}_{\mathrm{N}} & \mathrm{z}_{\mathrm{N}} \mathrm{x}_{\mathrm{N}} & \mathrm{x}_{\mathrm{N}} & \mathrm{y}_{\mathrm{N}} & \mathrm{z}_{\mathrm{N}} & 1
\end{array}\right]
\end{equation} is obtained from 3D points $(x_i,y_i,z_i)$ \cite{Ying2012}. This leads to the generalized eigenvalue problem
\begin{equation}
\mathbf{S}\mathbf{a}=\lambda\mathbf{M}\mathbf{a} 
\end{equation}
where $\mathbf{S}=\mathbf{D}^T\mathbf{D}$ \cite{Ying2012}. The potential ellipsoidal solution is the generalized eigenvector corresponding to the single positive generalized eigenvalue of $S$ and $M$ \cite{Ying2012}. For more details we refer to \cite{Ying2012}. From the obtained solution $\mathbf{a}$ one can construct the matrices 
\begin{equation}
\mathbf{P}=\left[\begin{array}{cccc}
\mathrm{a} & \mathrm{d} / 2 & \mathrm{f} / 2 & \mathrm{~g} / 2 \\
\mathrm{~d} / 2 & \mathrm{~b} & \mathrm{e} / 2 & \mathrm{~h} / 2 \\
\mathrm{f} / 2 & \mathrm{e} / 2 & \mathrm{c} & \mathrm{i} / 2 \\
\mathrm{~g} / 2 & \mathrm{~h} / 2 & \mathrm{i} / 2 & \mathrm{j}
\end{array}\right], \mathbf{Q}=\left[\begin{array}{ccc}
\mathrm{a} & \mathrm{d} / 2 & \mathrm{f} / 2 \\
\mathrm{~d} / 2 & \mathrm{~b} & \mathrm{e} / 2 \\
\mathrm{f} / 2 & \mathrm{e} / 2 & \mathrm{c}
\end{array}\right]
\end{equation}   
The center of the ellipsoid $\mathbf{x}_0$ is then obtained by solving 
\begin{equation}
\mathbf{Q}\mathbf{x}=\begin{pmatrix}-g\\-h\\-i\end{pmatrix}
\end{equation}
using a SVD decomposition. The radii of the ellipsoid can then be calculated 
by calculating the eigenvalues of $\mathbf{Q}$ translated to the obtained center $\mathbf{x}_0$ \cite{fitting_elipsoid}.

\begin{figure}
\includegraphics[width=\textwidth]{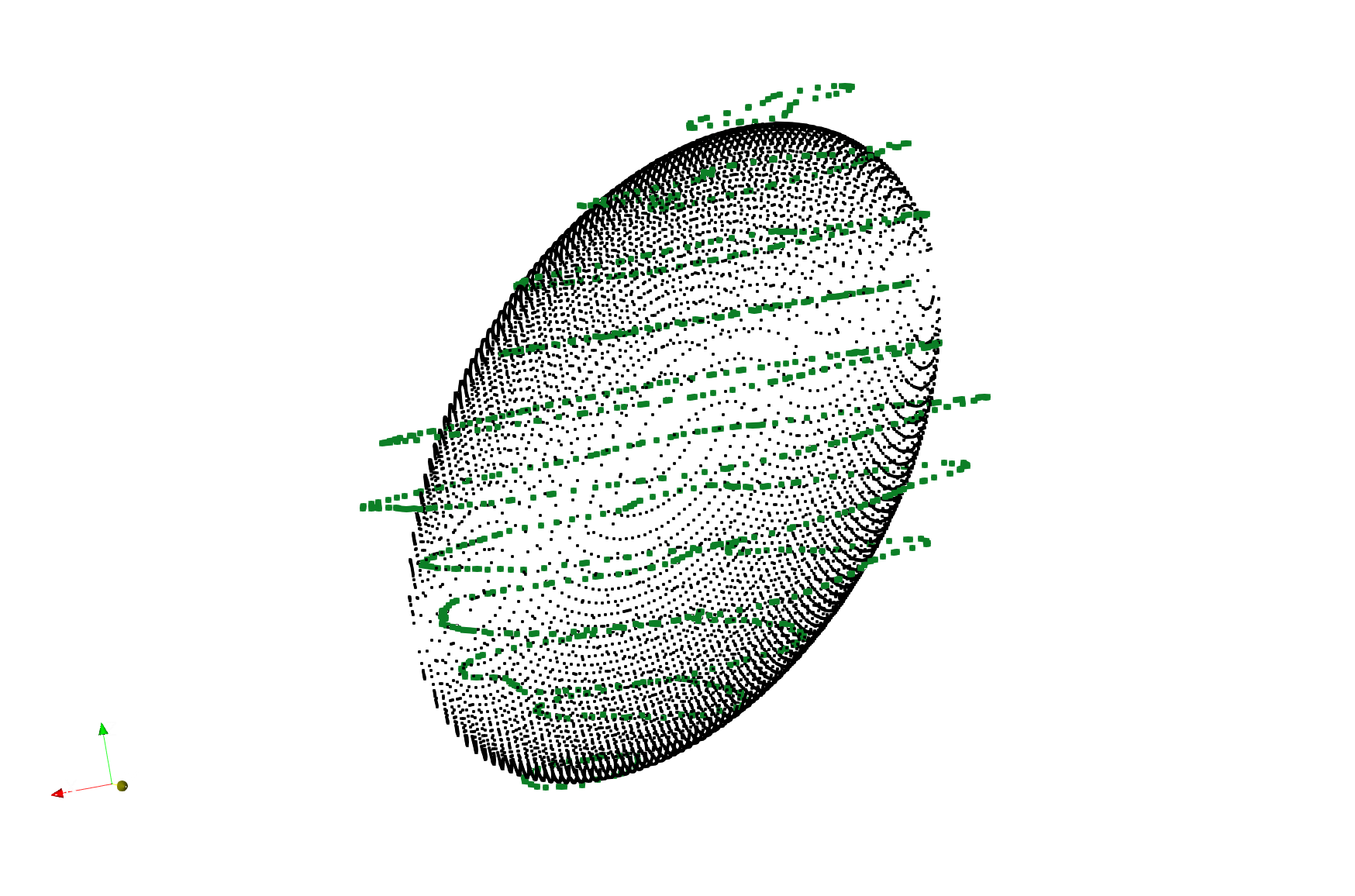}
\caption{Illustration of the ellipsoid fitting: GTV contour points are shown in green, points of fitted ellipsoid are shown in black}
\label{fig:illu_ellips_3d}
\end{figure}

\bibliographystyle{unsrtnat}
\bibliography{literatur}

\end{document}